\documentclass[12pt]{article}
\usepackage{epsfig, graphics}

\usepackage{subcaption}

\graphicspath{/Users/Joann/Desktop/COVID}

\usepackage{amsfonts}
\usepackage[latin1]{inputenc}
\newcommand{\nin}{\noindent}

\def\bkR{{\rm I\kern-.17em R}}

\def \1n{1\hskip -3pt \mbox{N}}

\def \Sum {\displaystyle \sum }

\newfont{\bbf}{cmbx12 scaled 1435}

\newcommand{\be}{\begin{equation}}
\newcommand{\ee}{\end{equation}}

\newcommand{\ba}{\begin{eqnarray}}
\newcommand{\ea}{\end{eqnarray}}
\newcommand{\ban}{\begin{eqnarray*}}
\newcommand{\ean}{\end{eqnarray*}}

\newcommand{\bt}{\begin{tabular}}
\newcommand{\et}{\end{tabular}}
\newcommand{\btb}{\begin{tabbing}}
\newcommand{\etb}{\end{tabbing}}
\newcommand{\bfr}{\begin{flushright}}
\newcommand{\efr}{\end{flushright}}

\newcommand{\bge}{\begin{enumerate}}
\newcommand{\ene}{\end{enumerate}}
\newcommand{\bc}{\begin{center}}
\newcommand{\ec}{\end{center}}

\def \1n{1\hskip -3pt \mbox{N}}

\begin{document}
\setlength{\baselineskip}{.26in}
\thispagestyle{empty}
\renewcommand{\thefootnote}{\fnsymbol{footnote}}
\vspace*{0cm}
\begin{center}

\setlength{\baselineskip}{.32in}
{\bbf Time Varying Markov Process with Partially Observed Aggregate Data; An Application to Coronavirus}\\

\setlength{\baselineskip}{.26in}
\vspace{0.5in}
C. Gourieroux \footnote[1]{University of Toronto,  Toulouse School of Economics and CREST,{\it e-mail}:
{\tt gouriero@ensae.fr}}, J. Jasiak  \footnote[3]{York University, Canada, {\it e-mail}:
{\tt jasiakj@yorku.ca},\\
The authors gratefully acknowledge financial support of the chair ACPR: Regulation and Systemic Risks, the ERC DYSMOIA, the ANR (Agence Nationale de la Recherche) and Natural Sciences and Engineering Research Council of Canada (NSERC)  \\
The authors thank A. Djogbenou, C. Dobronyi, Y. Lu, A. Monfort, P. Rilstone and J. Wu for helpful comments}
\vspace{0.1in}\\

Preliminary version, comments welcome 
\vspace{0.5in}
  
first version: March 31, 2020\\
revised: \today\\

\vspace{0.3in}
\begin{minipage}[t]{12cm}
\small

A major difficulty in the analysis of propagation of the coronavirus is that many infected individuals show no symptoms of Covid-19. This implies a lack of information on the total counts of infected individuals and of recovered and immunized individuals. In this paper, we consider parametric time varying Markov processes of Coronavirus propagation and show how to estimate the model parameters and approximate the unobserved counts from daily numbers of infected and detected individuals and total daily death counts. This model-based approach is illustrated in an application to French data.

\medskip

{\bf Keywords:} Markov Process, Partial Observability, Information Recovery, Estimating Equations, SIR Model, Coronavirus, Infection Rate.
\end{minipage}

\end{center}
\renewcommand{\thefootnote}{\arabic{footnote}}
\newpage

\newpage

\setcounter{page}{1}

\section{Introduction}

The aim of this paper is to address the problem of partial observability, encountered recently in epidemiological research on Covid-19. More specifically, some individuals are infected and asymptomatic. Therefore, they remain undetected and not recorded \footnote{Even though some data on asymptomatic ratios are available [see e.g. Nishiura et al.(2020)],  some individuals may remain undetected for other reasons, such as refusing to be tested, or getting false negative tests results.}. As a consequence, the total count of recovered and immunized individuals is also unknown, as only the number of recovered and detected individuals is available. This problem of partial observability of counts renders difficult the estimation of an epidemiological SIRD (Susceptible, Infected, Recovered, Deceased) model, extended to disentangle the infected and undetected from the infected and detected individuals. Moreover, such substantial undocumented infection can facilitate the rapid propagation of the virus (Li et al.(2020)).

There are two solutions to this difficulty: The first one is to sample the population daily and perform serological tests on those samples in order to estimate the proportions of infected and undetected and of recovered individuals. The current epidemic shows that  validating and producing reliable serological tests can take time. Moreover, regularly performed samplings can be costly especially in terms of time of health care providers. The second approach, proposed in this paper, is purely model-based. Loosely speaking, under the standard extended SIRD model, the evolution of death rates might be different, depending on whether all infected individuals are detected or not. This implied difference will  allow us for a model-based estimation of the proportions of infected undetected individuals (resp. recovered immunized) [see, Verity et al.(2019) for pure model based estimation of coronavirus infection,  Manski,Molinari (2020) for set estimation of the infection rate].

The paper discusses the general case of time varying Markov processes when aggregate counts are partially observed. Section 2 describes the latent model of qualitative individual histories. These histories follow a time varying Markov process with transition probabilities that can depend on latent counts and unknown parameters. The observations are functions of the frequencies of the different states (compartments in epidemiological terminology), although not all of those frequencies are observed, in general. More specifically, only some states can be observed and/or a sum of frequencies over subsets of states can be observed. Section 3 introduces the estimation method, which jointly estimates the unknown parameters and the unknown state probabilities. We derive the asymptotic properties of the estimators under identification. Identification, which is the main challenge of the proposed approach is the topic of Section 4. First, we discuss the identification in a homogeneous Markov, i.e. when the transition matrix is not time varying. Without additional restrictions on the transition probabilities, that model is not identifiable and the proposed approach cannot be used. However, it is not the case for a time varying Markov process that includes contagion effects and, and in particular for the SIR-type models used in epidemiology. The estimation approach is illustrated in Section 5 with a SIR type model for French data. Section 6 concludes. Some technical  problems are discussed in the Appendices.

\section{Latent Model and Observations}

\subsection{Latent Model}

We consider a large panel of individual histories $Y_{i,t}, i=1,...,N,\,t=1,..,T$,where the latent variable is qualitative polytomous with $J$ alternatives denoted by $j=1,..,J$.

\medskip
\nin {\bf Assumption A1:} The individual histories are such that:

\nin i) The variables $Y_{i,t}, \; i=1,..,N$, at $t$ fixed, have the same marginal distributions. This common marginal distribution is discrete and summarized by  the $J$-dimensional vector $p(t)$, with components:

$$p_j(t)= P(Y_{i,t}=j).$$

ii) The processes $\{Y_{i,t}, \,t=1,..,T\}, \;i=1,..,N$ are independent (heterogeneous) Markov processes with transitions from date $t-1$ to date $t$ summarized by a $J \times J$ transition matrix  $P[ p(t-1); \theta]$, parametrized by $\theta$. This matrix is such that each row sums up to 1.

\medskip
Thus, we consider a discrete time model applicable to data on a homogeneous population of risks.
Let $f(t)$ denote the cross sectional frequency, i.e. the sample counterpart of $p(t)$. It follows from the standard limit theorems that:

\medskip
\nin {\bf Proposition 1:} Under Assumption A1 ,the $f(t)$'s are asymptotically normal for large $N$. 

\medskip

This specification of the transition matrix includes the homogeneous Markov chain, when there is no effect of lagged $p(t-1)$. It also includes the standard contagion models of SIR type used in epidemiology [see, McKendrick (1926), Kermack, McKendrick (1927) for early articles on SI and SIR models in the literature, Hethcote (2000), Brauer et al.(2001), Vinnicky, White (2010) for general presentations of epidemiological models, Allen (1994) for their discrete time counterparts, Gourieroux, Jasiak (2020) for an overview, and also examples given below].

\nin As vectors $p(t)$ change over time, stationarity is not assumed.

\subsection{Observations}

In practice, the individual histories may not be observed, while cross-sectional frequencies are available. These can be the frequencies $f(t),t=1,...,T$, or aggregates of such frequencies.
\medskip

\nin {\bf Assumption A2:} The observations are: $\hat{A}_t = A f(t),t=1,..,T$, where $A$ is a $K \times J$ state aggregation matrix, that is a matrix with rows containing zeros and ones. The aggregation matrix is known and of full rank K.

\medskip

\nin {\bf Example 1:} When $A=Id$, all $f(t)$'s are observed. This is the case considered in McRae (1977), Miller, Judge (2015).

\medskip
\nin {\bf Example 2:} In a model of the coronavirus propagation, the following 5 individual states can be distinguished:  $1=S$, for Susceptible, $2=IU$, for Infected and Undetected, $3=ID$ for Infected and Detected, $4=R$ for Recovered, and $5=D$ for Deceased. The observed frequences are $f_3(t)$ and $f_5(t)$, but not the other frequencies. We have a $2 \times 5$ matrix $A$  given by:

$$ A =
\left[ \begin{array}{ccccc} 
0 & 0 & 1 & 0 & 0 \\  
0 & 0 & 0 & 0 & 1
\end{array} \right],$$

\nin that characterizes the selection of the frequencies.

\medskip

\nin {\bf Example 3:} In some other applications, matrix $A$ truly aggregates the frequencies, as, for instance, in cascade processes and percolation theory applied to an epidemiological model \footnote{See Good (1949) and Hammersley (1957) for the introduction of cascade processes and percolation, resp., in the literature.}. Let us consider a country with two regions and a SI model distinguishing these regions. We get a 4 state model: 1=S1, susceptible in region 1, 2=S2, susceptible in region 2, 3=I1, infected in region 1, 4=I2, infected in region 2. A propagation model can be written at a disaggregate level to account for both propagation within and between regions. Thus, there is a competition between the contagions coming from regions 1 and 2. However, only aggregate data for the entire country may be available. Then, the aggregating matrix A is equal to:

$$ A= \left[\begin{array}{cccc} 1 & 1 & 0 & 0 \\ 0 & 0 & 1 & 1 \end{array} \right] .
$$

Although the process of aggregate counts: $f_1(t) + f_2(t), f_3(t) + f_4(t)$ may not be Markov, in general, it is important to consider the special case when it is, and then explore the possibility of identifying the parameters of regional, i.e. disaggregated dynamics. This is exactly the objective of percolation theory [see, Garet, Marchand (2006) for a detailed analysis of competing contagion sources].

\section{Estimation}

Under Assumption A1, we can use the Bayes formula to relate the marginal theoretical probabilities $p(t)$ to the transition probabilities by:

\begin{equation}
p(t)= P[p(t-1); \; \theta]' p(t-1), \; t=2,...,T.
\end{equation}

\nin The nonlinear recursive equation (3.1) is the discrete time counterpart of the deterministic differential system commonly used in epidemiology [see, Gourieroux, Jasiak(2020)]. These equations will be used as the estimating equations in the asymptotic least squares estimation method outlined below \footnote{see, Godambe, Thompson (1974),Hardin, Hilbe(2003)}.
In our framework, the parameter of interest includes $\theta$ as well as the sequence of vectors $p(t)$. They can be jointly estimated from the following optimization:

$$(\hat{p}(1),..,\hat{p}(T), \hat{\theta})
= ArgMin \Sum_{t=2}^T || p(t) - P[p(t-1);\theta]'p(t-1) || ^2$$

$$\mbox{s.t.} \;\; A p(t)= A f(t)=\hat{A}_t, t=1,..,T,$$

\nin where $||.||$ denotes the Euclidean norm.

\medskip
\nin {\bf Proposition 2:} If the constrained optimization given above has a unique solution that is continuously differentiable with respect to $A f(t)= \hat{A}_t, t=1,...,T$, then the estimator is asymptotically consistent, converges at rate $1/\sqrt{N}$, and is asymptotically normally distributed.

\medskip

\nin The expression of the asymptotic variance-covariance matrix is derived in Appendix 2.

If $A=Id$, that is, if all frequencies are observed, we obtain the case analysed in McRae (1977). In the general framework, this optimization is not only used to estimate parameter    $\theta$, but also to approximate the unobserved marginal probabilities.

The condition given in Proposition 2 is an identification condition, which is discussed in detail in the next section.

\section{Identification Condition}

In this section we discuss the (asymptotic) identification corresponding to the objective function given in Section 3. This objective function has a simple form, as it is quadratic under linear constraints with respect to the sequence $p(t)$. This allows us for an optimisation in two steps: first with respect to the $p(t)$'s, and next, with respect to $\theta$ after concentrating. This is the approach used below for identification \footnote{This numerical simplification will not arise for other measures of distance in the Cressie-Read family between probability distributions  [Cressie,Read (1984), Miller,Judge (2015)].}.

\subsection{Order Condition}

By taking into account the fact that probabilities sum up to one, we can compare the number of moment conditions equal to $ (J-1)(T-1)$ with the number of parameters of interest that is $(J-K-1) T$ + $\dim \theta$,  iff $KT \geq \dim \theta$. Therefore the order condition is satisfied iff the number of days $T$ is sufficiently large. Let us now consider the rank condition.

\subsection{Rank condition for Homogenous Markov}

For ease of exposition, we consider a homogenous Markov model with 3 states: $J=3$. The parameter $\theta$ includes the elements of the transition matrix $P$, which has 6 independent components, given that each row of $P$ sums up to 1. We assume that the observed marginal probabilities are $p_3(t),t=1,..,T$. The Bayes formula:

\begin{equation}
p(t)= P' p(t-1), \; t=2,...,T,
\end{equation}

\nin leads to $2(T-1)$ independent moment restrictions that are the estimating equations: 

\begin{eqnarray*}
p_2(t) & = & p_{12} p_1(t-1) +p_{22} p_2(t-1) + p_{32} p_3(t-1) \\
p_3(t) & =  & p_{13} p_1(t-1) + p_{23} p_2(t-1) + p_{33} p_3(t-1),
\end{eqnarray*}

\nin or equivalently,

\begin{eqnarray}
p_2(t) & = & p_{12} [1-p_2(t-1)-p_3(t-1)] + p_{22} p_2(t-1) +p_{32} p_3(t-1) \nonumber\\
p_3(t) & = & p_{13} [ 1-p_2(t-1)-p_3(t-1)] + p_{23} p_2(t-1) + p_{33} p_3(t-1)  
\end{eqnarray}

\nin To discuss identification, we look for the solutions in $P$ and $p(t),t=1,...,T$ of system (4.2) written for $t=2,...,T$. We have the following result:

\medskip
\nin {\bf Proposition 3:} Generically, i.e. up to a (Lebesgue) negligible set of parameter values,and if $T \geq 6$, we have that:

i) Parameter $P$ is not identifiable, with an under-identification order equal to 3.

ii) There exist 3 functions of $P$ that are identifiable. These functions are independent of $T$.

iii) These functions are over-identified with an over-identification order equal to $T-5$.

\medskip
\nin {\bf Proof:} The proof is based on a concentration with respect to the values of $p_2(t)$. From the second equation of system (4.3), we see that $p_2(t-1)$ is a linear affine function of $p_3(t), p_3(t-1)$, with coefficients that depend on $P$. These linear affine expressions can be substituted into the first equation of system (4.3) to show that the observed sequence $p_3(t)$ satisfies a linear affine recursion of order 2:

$$ p_3(t)= a(P) + b(P) p_3(t-1) + c(P) p_3(t-2),\; t=3,...,T,$$

\nin with coefficients that depend on $P$. The results follow since:

i) the functions $a(P), b(P), c(P)$ are identifiable;

ii) the degree of under-identification of $P$ is: 6-3=3;

iii) the degree of over-identification of the identifiable parameters is: $T-2-3=T-5$.

                                                                                                                            \hfill Q.E.D.

\medskip
Appendix 3 provides the expressions of functions $a(P), b(P), c(P)$ and points out that Proposition 3 does not apply only under circumstances that are negligible. In particular, identification requires that observations $p_3(t)$ correspond to a nonstationary episode as shown in the remark below.
\medskip

\nin {\it Remark 1:} Let $\pi$ denote the stationary probability solution of the Markov chain, defined by:

$$ \pi = P' \pi.$$

\nin If the observed $p_3(t)=\pi_3$ were associated to a stationary episode, the only identifiable function of parameters would be $\pi_3(P)$ and the under-identification degree would be equal to 6-1=5. Therefore, observing the process during a nonstationary episode provides a gain of 2 identification degrees.
\medskip

\nin {\it Remark 2:} If the Markov structure is recursive, that is, if matrix $P$ is upper triangular, the under-identification degree becomes 3-3=0, and the parameter is generically identifiable.

\medskip
Proposition 3 shows that  we can expect to identify the parameter of interest if we either consider a) a homogeneous Markov and constrain the parameters as illustrated in Remark 2 by an example of the recursive system,, or b) a non-homogeneous Markov discussed in the next subsection.

\subsection{Rank Condition in a Propagation Model}

Let us now consider an epidemiological model with $J=3$ states  to facilitate the comparison with Section 4.2. The states are: 1=S for susceptible, 2=I for infectious (individuals stay infectious, even if they recover), 3=D for deceased. The rows of the transition matrix are the following:

row $1=S: \;\; (1-p_{13}) [1-logist(a_1 + a_2 p_2(t-1))]; \;\;(1-p_{13}) logist (a_1 + a_2 p_2(t-1)); \;\; p_{13} $

row $2=I:\;\; 0; \;\; 1-p_{23}; \;\; p_{23} $

row $3=D: 0, \;\; 0, \;\; 1$

\nin where $logist(x)= 1/[1+exp(-x)]$ is the logistic function, i.e. the inverse of the logit function. We obtain a triangular transition matrix with state D as an absorbing state. The contagion effect is characterized by parameter $a_2$ and follows a nonlinear logistic function. We also expect that mortality rate $p_{23}$ is strictly larger than mortality rate $p_{13}$. There are 4 independent parameters in $\theta = [ a_1, a_2, p_{13}, p_{23}]'$.

\medskip
\nin {\bf Proposition 4:} The SID model given above is generically identifiable. Parameter $\theta$ is over-identified with an over-identification order equal to 5.

\medskip
\nin {\bf Proof:} The proof is similar to the proof of Proposition 3. The two independent moment conditions are:

\vspace{-0.1in}

\begin{eqnarray*}
p_2(t) &= & (1-p_{13}) logist[ a_1+a_2 p_2(t-1)] [1-p_2(t-1)- p_3(t-1)] +(1-p_{23}) p_2(t-1) \\
p_3(t) & = & p_{13}[1-p_2(t-1)-p_3(t-1)] + p_{23} p_2(t-1).
\end{eqnarray*}

\nin From the second equation, it follows that $p_2(t-1)$ is a linear affine function of
$p_3(t)$ and $p_3(t-1)$. Next by substituting into  the first equation, we find that the observed $p_3(t)$ satisfies a nonlinear recursive equation of order 2 of the type:

$p_3(t)= a_1(\theta) + b_1(\theta) p_3(t-1) + c_1(\theta) p_3(t-2)
+[a_2(\theta) + b_2(\theta) p_3(t-1) + c_2(\theta) p_3(t-2)]\, logist[ a_3(\theta) + b_3(\theta) p_3(t-1) + c_3(\theta) p_3(t-2)].$

\medskip

\nin If $T$ is sufficiently large, this nonlinear observed dynamics allows us to identify 9 nonlinear functions of parameter $\theta$. Thus parameter $\theta$ is identifiable with an over-identification order equal to 5.

                                                                                                                        \hfill Q.E.D.

\medskip

From Remark 2, we expected that the triangular form of the transition matrix alone would facilitate the identification. However, the order of over-identification reveals the additional role of the contagion effect. The nonlinear dynamics induced by the logistic transformation also facilitates identification.

\section{ An Illustration}

This section illustrates the estimation approach and its performance in an epidemiological model.

\subsection{The model and the observations}

We consider a model with  5 states:
1=S, 2=IU, 3=ID, 4=R ,5=D, and the following rows of the transition matrix:

\medskip
\nin row 1: $(1-p_{15}) \pi_{11t} ; \; (1-p_{15}) \pi_{12t} ; \; (1-p_{15}) \pi_{13t}; \;0; \; p_{15}$

where the $\pi_{1jt},  j=1,2,3 $ sum up to 1, and are proportional to:

\nin $\pi_{11t} \approx 1; \; \pi_{12t} \approx exp(a_1+b_1 p_2(t-1) + c_1 p_3(t-1)]; \;\pi_{13t} \approx exp[a_2 + b_2 p_2(t-1) +c_2 p_3(t-1)]$

\nin row2: $0; p_{22}; p_{23}; p_{24}; p_{25}$

\nin  row3: $0; 0; p_{33}; p_{34}; p_{35}$

\nin  row 4: $0;0; 0; p_{44}; p_{45}$

\nin row 5: $0; 0; 0; 0; 1$

Conditional on staying alive, the first row includes a multinomial logit model for the competing propagation driven by  either lagged IU, or lagged ID [see, e.g. McFadden (1984)]. The propagation parameters allow for different impacts of $p_2(t-1)$ and $p_3(t-1)$, since the detected individuals are expected to be self-isolated more often. There is no contagion effect from the recovered R, who are assumed no longer infectious \footnote{For viruses other then Covid-19, the recovered, immunized individuals can stay infectious.}. The structure of zeros in the transition matrix indicates that one cannot recover without being infected, one cannot be infected twice (as expected for COVID 19) and death is considered as an absorbing state.

\medskip
This is a parametric model with 6+7=13 parameters, that are the 6 parameters $a_l, b_l, c_l,
l=1,2$ and 7 independent transition probabilities.

Among the 5 series of frequencies $f_j(t)\; j=1,..,5$ that sum up to 1 at each date, $f_3(t)$ and $f_5(t)$ of infected detected and of deceased, respectively, are assumed to be observed. The frequencies
$f_2(t)$ and $f_4(t)$ are unobserved and will be considered as additional quantities of interest to be estimated jointly. They are crucial for a model-based reconstituting of counts of infected undetected and of recovered immunized individuals.

As illustrated in Section 4.3,  the triangular form of the transition matrix and the nonlinear doubly logistic contagion model provide  generic identification.

\subsection{Simulations}

The model above can be used for simulating the Covid-19 propagation for given values of parameter $\theta$ and initial value $p(1)$. These values are set as follows:

The daily mortality rates are: $p_{15}=p_{45}= 3e-05$, $p_{25}=0.004, p_{35}=0.013$. The mortality rates $p_{15}=p_{45}$ correspond to the long term mortality rates in France;
$p_{35}$ is an average mortality rate of individuals detected with Covid-19 in hospitals [see, Verity et al.(2020), Table 1 for a comparison], $p_{35}$ has been fixed between those numbers to account for a lower rate due to the presence of asymptomatic individuals [see e.g. Nishiura et al.(2020) for the asymptomatic ratio].

We assume, there are about 3 times more transition to IU than to ID,i.e.

$$ \exp(a_1) = 3 \exp(a_2),\; b_1=b_2, \;c_1=c_2,$$

\nin and the propagation effects due to IU and ID, are equal i.e. $b_2=c_2$. Then $a_2, b_2$ are set such that:

$ \exp(a_2)=1e-06$ and $\exp(2 b_2/1000)=25$. These parameters have been set to provide about 60 new daily detected infections at the beginning of the epidemic for a population of 60 millions of inhabitants and of 1500 new daily infected later on, about 30 days after the beginning.

\nin The parameters $p_{23}, p_{24}, p_{34}$ are set equal to:

$p_{24}=p_{34}=0.03$, representing an average recovery time of about 33 days before being immunized. This average time is fixed equal for the IU and ID states in the simulation.

Rate $p_{23}$ is fixed equal to $p_{12}=1e-06$, which is the detection rate $\exp(a_2)$. 

Coefficient $a_2$ is strictly positive. This means that there can exist exogenous sources of infections for the population of interest, either from animals to humans, or more importantly from humans of another population to humans in the population of interest, due to tourism and migration. We consider an open economy from the epidemiological point of view \footnote{The idea of collective immunity that implies, the infection will disappear if more than 60\% of people are immune, implicitly assumes a closed economy. It is valid for the world in its entity, but not for each country separately.}.
We do not account for the increasing effort of performing daily tests during the epidemic (its outcome has not been substantial in France during this period, due to the pressure on the market of test components \footnote{Our model does not explicitly take into account the reliability of these tests, i.e. the proportion of false negative outcomes.}).

Next, the  parameters corresponding to the diagonal transition probabilities are computed from the unit mass restriction on each row.

\nin All probabilities of transitions out of the diagonal are all very small as a consequence of the daily frequency of our data. The initial marginal probabilities are set equal to:
$p_0(t)=(1,0,0,0,0)$, that corresponds to an initial population with no prior infection from the coronavirus in this population. Thus, the first cluster will be due to travellers entering the country.

Two types of dynamic analysis can be performed, depending whether the sequence of $p(t)$, or the sequence of $f(t)$ are considered. The dynamics of $p(t)$'s are deterministic, and driven by the deterministic system (3.1). They provide us the dynamics of the expected values of $f(t)$'s. The dynamics of $f(t)$'s are stochastic with trajectories obtained by simulating the time varying Markov process. As an additional outcome, the difference between the $p(t)$'s and $f(t)$'s provides a measure of  uncertainty on any predictions obtained from the deterministic model of $p(t)$'s.

Figure 1 shows the evolutions of $p_2(t), p_3(t), p_4(t), p_5(t)$ in separate panels as their ranges and evolutions differ, due to the selected parameter values. In addition, Figure 1 illustrates the effect of an increase (decrease) of propagation parameter values $b_1, b_2, c_1$ and $c_2$ on the marginal probabilities.  

\medskip
 [Insert Figure 1: Evolutions of Marginal Probabilities]
\medskip

\nin The solid lines represent trajectories of $p(t)'s$ computed from the baseline parameter values given above. The  dotted and dashed lines, respectively, depict the trajectories obtained when parameters $b_1, b_2, c_1$ and $c_2$ increase and decrease by a factor of 2, respectively.

The change of propagation parameters,  has an impact on the shape of curves, resulting in faster (slower) rates of increase in all panels, except for the bottom right one. The dynamic of $p_5(t)$ does not seem affected, as the trajectories computed from baseline and increased (decreased) parameter values overlap one another. 

Figure 2 displays the evolutions of $p_3(t)-p_3(t-1)$ and $p_5(t)-p_5(t-1)$ multiplied by the total size of the population, i.e. 60 millions. These are the new counts of ID, to be compared with the health system capacity, and the numbers of new deaths D, including, but not limited to the confirmed deaths from Covid-19.

\medskip
[Insert Figure 2: Evolution of New Counts]
\medskip

\nin As before, the solid lines represent the trajectories computed from the baseline parameter values and the dotted and dashed lines show the trajectories obtained by increasing and decreasing the parameter values, respectively. A change in propagation parameters affects the shape of the curves of new counts, resulting in higher (lower) growth rates of new counts.
 
The evolutions are computed over a period of 60 days, i.e. 2 months. During this episode, the total number of infected individuals remains rather small, as compared to the size of the population and so does the total count of deaths. The above figures have to be interpreted in terms of stocks and flows as the numbers associated with R and D (resp. S) are cumulated and are interpretable as stocks. This cumulation effect explains the increasing patterns in Figure 1, with higher rates for higher values of propagation parameters.

The two Infected states IU and ID are flows, as they are observed between the times of entry in,  and exit from the state of infection. Moreover, the probability of exiting after 20 days is very close to 1. We usually expect a "phase" transition effect:
For small $t$, these counts increase quickly as new infected individuals are cumulated without a sufficiently high number of exits to compensate for the arrivals. This explain an increase  of the curves at the beginning of the period. After that initial period, the counts of  exits tend to grow and offset the new arrivals so that the curves tend to flatten. More precisely, they continue to increase, due to the propagation effect, but at a very low rate. This is the so-called flattening of the curve. 
This theoretical evolution depends on the choice of parameter values, especially the propagation parameters. Given the selected parameter values that allow for exogenous sources of infection, the initial convex pattern in the counts of infected is not visible. Only the concave part of the curve, up to its flat part, is observed. One can perform similar dynamic analysis for other credible scenarios, which is called the sensitivity analysis.

The Figures given above have been simulated with time independent propagation parameters. A self-isolation measure introduced at some point, would have changed  subsequent evolution. There is  a first a tendency to reach a flat part on the curve without self-isolation, and then to reach a lower flat part on the curve with self-isolation measures. Therefore over a longer period, the first flat part can appear as a smoothed peak. If self-isolation measures are lifted afterwards, a second peak of infections is expected, and so on, resulting in a sequence of stop and go [Ferguson et al.(2020)].

\subsection{Estimation}

This section presents the estimation of the extended SIRD model from data on Covid 19 propagation in France over the period of 22 days between 03/16  to 04/06, 2020.

The model introduced in Section 5.1 assumes a stable environment of constant social distancing measures, which was the case in France during the observation period. A total lock-down  was implemented on the weekend of March 16, with the closure of shops, schools, universities and introduction of strict social distancing rules. This self-isolation measure had an impact on the spread of the disease,  especially on the propagation parameters and some mortality parameters \footnote{The effect of Covid-19 on the total mortality rate is  unclear. There is a negative effect of the virus, but also positive effect due to better protection against other viruses such  as influenza, and reduced number of car accidents.}. To detect that effect, it would be necessary to estimate separately the model over the periods of March  1 to 15, and  March 20 up  to April 7, which is possible as these sufficiently long for identification (see Proposition 3) \footnote{It is not the case for countries where the propagation is too recent, or isolation decided too late, or data unreliable at the beginning of the episode (Wuhan), or the isolation period too short 
 (Denmark) or introduced in successive steps, or isolation decisions being very different depending on the regions of (Germany and the US).}. Then, we could  compare the results to measure the efficiency of the lock-down and perform predictions including the effects of different stages of reopening.
 
However we focus on data that are reliable for estimation purpose. The fully observed states are the states ID and D. State ID is assumed equivalent to hospitalization, as the counts of ("confirmed") detected, which are publicly available, are measured with error. This is due to the counts of detected individuals being derived from test results, while not all tests results may have been recorded, some people could have been tested multiple times, inflating the counts, or people have not been  tested at random or without an adequate exogenous stratification, which creates selectivity bias \footnote{Similar data are used for estimation in Manski ,Molinari(2020), but not adjusted for the significant selectivity bias.}.
In contrast, the hospitalization data are reliable and regularly updated.
State D is assumed observed through total death counts. It includes death from Covid-19, which are reported on-line as D/H, i.e. death after hospitalization and known to underestimate the true D/H values true number of deaths due tu coronavirus, as these do not include all deaths from COVID-19 at home, or in long term care homes.

The series to be estimated are infected undetected IU and recovered R. We use the available series of ("confirmed") detected and of recovered after hospitalization for comparison with the estimates.

More specifically, we use the French data on the total daily number of deaths from the French National Statistical Institute INSEE( 2020) and the daily data on coronavirus pandemic from Sante Publique France (2020) reported at https://dashboard.covid19.data.gouv.fr/ and
https://www.linternaute.com/actualite/guide-vie-quotidienne/2489651-covid-19-en-france-les-dernieres-statistiques-au-06-avril-2020/ \footnote{The size of the French population is 66,9 millions of inhabitants.}. The daily evolutions of total counts of hospitalized, detected, recovered and deceased individuals reported in those sources are displayed in Figure 3.

\medskip
[Insert Figure 3: Evolution of Observed Counts, 03/16 to 04/06, France]

\medskip
The panels display the series of  "hospitalized", "confirmed" (i.e. detected), "returned from hospital" (i.e. recovered after hospitalization) in the top row and left bottom panels, respectively.
In the bottom right panel, the dynamics of counts of total deceased (solid line) and deceased due to Covid-19 (dashed line) are distinguished. 

The estimation of the model introduced in Section 5.1 produced the following results.
The estimated coefficients are  $a_1= -8.6517, 
 a_2 = -11.1481, 
 b_1 =  0.0034, 
 b_2 =  2.499e-05, 
 c_1 =  8.482e-05, 
 c_2 =  0.00028$. 
The estimated coefficient of mortality rate $p_{15}$ is 3.1575e-05, which is close to the  mortality rate in France of 3e-05 = 0.03/1000, used in the simulation study in Section 5.2.


The estimated transition matrix is given below in Table 1:

\newpage
\begin{center}

Table 1. Estimated Transition Matrix
\end{center}

\begin{center} 
\begin{tabular}{|c|c|c|c|c|c|}
\hline

    & 1= S & 2=IU  & 3=ID & 4=R & 5=D   \\ \hline
2=IU & 0 & 0.9022 & 0.0386 & 0.0571 & 0.00207 \\ \hline
3=ID & 0 & 0 &  0.7926 & 0.1032 &  0.0158 \\ \hline
4=R & 0 & 0 &  0 & 0.9999   &    1.514e-5 \\ \hline
5=D & 0 & 0 & 0 & 0 & 1 \\ \hline

\end{tabular}

\end{center}

\nin The evolutions of estimated counts of IU, i.e. infected and undetected and of R, i.e. recovered are shown in Figure 4 (solid line). The estimates are compared with the available counts of ("confirmed") detected individuals and of 
recovered after being hospitalized (R|H).

\medskip

[Insert Figure 4: Estimated and Observed Counts]

\medskip
The estimated counts exceed those reported by the sources. In particular,
the observed and estimated counts on April 06, 2020, which is the last day of sample are as follows:
The final observed count of ("confirmed") detected  is equal to 78167 and is 1.2 times smaller that the  estimated final count of infected and undetected (IU) equal to 94461.
The observed final count of Recovered (after being hospitalized) equal to 17250 is 6.24 times smaller than the estimated final count of Recovered equal to 107640.

Let us now present a scenario of a projected evolution, based on the estimated coefficients values and
probabilities. 
Figure 5 below shows the projected evolution of the marginal probabilities 
of IU, ID, R and D over the period of 25 years.

\medskip

[Insert Figure 5: Projected Evolution of Marginal Probabilities]

\medskip

\nin Figure 5 displays peaks in marginal probabilities of states 2 and 3 that occur after about 98 days. At the peak, the  projected count of infected and undetected (IU) individuals is over 300,000. Another pessimistic outcome is that it takes about 25 years for the marginal probabilities to decline to 0. Figure 6 below shows the projected daily new counts  of ID and deceased (D)
over an initial period of 60 days, which can be used for assessment of the capacity of the health sector.

\medskip
[Insert Figure 6: Projected Evolution of New Counts]

\medskip

\nin We observe that after 60 days, the daily new death counts increase to 3000 per day. At the same time, the new hospitalizations decrease quickly over time.

\section{Concluding Remarks}
This paper is intended to provide a solution for incomplete counts of infected and  undetected  individuals and  of recovered individuals. These unknown quantities can be estimated jointly with the parameters in a standard framework of a compartmental epidemiological model. This approach is illustrated in an estimation involving French count data.  The methodology required daily data on the total number of deaths, not only on the deaths due to COVID. These data are available in France and other European countries [see, the website Euromomo], but are unavailable in other countries, such as Canada.

The estimation approach can be extended as follows, which is left for future research: i) The model is a special nonlinear state space model, with states $p(t)$, deterministic state equations (3.1), and  (Gaussian) measurement equations:
$\hat{A}_t =A p(t) + u(t)$, where $u(t)$ denotes the (approximately Gaussian) measurement error. Even though the joint estimation can be easy performed, as shown in Section 5.3, the general algorithms for nonlinear state space models would be preferred for updating the results at each daily arrival of new observation. For this purpose, a Robust Extended Kalman Filter (EKF) [Julian, Uhlman (1997), White (1999), Krener (2003)],an Unscented Kalman Filter (UKF) [Song,Grizzle (1995), Wan,Van der Merwe(2000)], and particle filters [Pitt, Shephard (1999), Schon et al.(2005)] can be applied.
ii) Additional unobserved state variables could also be introduced to account for individual compliance with self-isolation measures and its dynamic [see e.g. Alvarez et al.(2020), Chudik et al.(2020),Ferguson et al.( 2020), Tang et al. (2020)]. The individual efforts (moral hazard phenomenon) have impact on the propagation parameters. This can be captured by introducing propagation parameters with stochastic heterogeneity over time. Specific heterogeneity dynamics would allow for reproducing the stop and go phenomenon [see e.g. Ferguson et al.(2020), Figure 4]. iii) The treatment of missing data can likely be improved by introducing additional explanatory variables that are expected to impact the virus propagation. This approach is followed in Hortacsu et al. (2020) who use hospitalization data from various regions and  interregional transportation data to forecast infection rates. iv) Other specifications of the propagation functions $\pi_t$ can also be considered and compared [see Wu et al. (2020)].

\newpage

{\bf {\Large REFERENCES}} \\

\nin Allen, L.J.S. (1994): Some Discrete-Time SI, SIR  and SIS Epidemic Models, Mathematical Biosciences, 124, 83-105. \\

\nin Alvarez, F., Argente,D. and F.,Lippi (2020) :A Simple Planning Problem for COVID-19 Lockdown,DP University of Chicago.\\

\nin Brauer, F. and C. Castillo-Chavez (2001): Mathematical Models in Population Biology and Epidemiology,Springer,New York. \\

\nin Chudik, A., Pesaran, H., and A., Rebucci (2020): Voluntary and Mandatory Social Distancy: Evidence on COVID 19 Exposure Rates from Chinese and Selected Countries,NBER 27034. \\

\nin Cressie,N.  and T., Read (1984): Multinomial Goodness-of-Fit Tests, JRSS,B,46,440-464.\\

\nin Einicke,G. and L., White (1999): Robust Extended Kalman Filtering,IEEE Trans. Signal. Process.,47,2596-2599.\\

\nin Ferguson N., et al. (2020): Estimating the Number of Infections and the Impact of Non-Pharmaceutical Interventions on Covid-19 in 11 European Countries,Imperial College London.\\

\nin Garet,O., and I.,Marchand (2006): Competition Between Growths Governed by Bernoulli Percolation,Polymath,12,695-734. \\

\nin Godambe,V and M Thompson (1974) Estimating Equations in the Presence of a Nuisance Parameter,Annals of Statistics,3,568-571.\\

\nin Good I. (1949): The Number of Individuals in a Cascade Process, Proc. Camb. Phil. Soc.,45,360-363.\\

\nin Gourieroux,C., and J., Jasiak (2020): Analysis of Virus Propagation: Anatomy of Stochastic Epidemiological Models, DP.\\

\nin Hammersley,J. (1957): Percolation Processes: Lower Bounds for the Critical Probability, Annals Math. Stat.,28,790-795. \\

\nin Hardin,J and J Hilbe (2003): Generalized Estimating Equations, Chapman \& Hall \\

\nin Hethcote, H. (2000): The Mathematics of Infectious Diseases,SIAM Review, 42, 599-653. \\

\nin Hortacsu,A., Liu, J.,and T., Schwieg (2020):Estimating the Fraction of Unreported Infections in Epidemics with a Known Epicenter: An Application to COVID-19, University of Chicago DP.\\

\nin Julien,S and J., Uhlman (1997): A New Extension of the Kalman Filter to Nonlinear Systems,11th Int.Symp. on Aerospace/Defence,Sensing,Simulation and Controls\\

\nin INSEE (2020): Nombre de deces quotidiens par departement, April 10. \\

\nin Kermack, W., and A. McKendrick (1927): A Contribution to the Mathematical Theory of Epidemics,Proceedings of the Royal Statistical Society, A, 115,700-721. \\

\nin Krener,A (2003):"The convergence of the EKF"in Directions in Mathematical Systems:Theory and Optimization,173-182, Springer \\

\nin Li, R. et. al.(2020): Substantial Undocumented Infection Facilitates the Rapid Dissemination of Novel Coronavirus(SARs-CoV2), Science. \\

Manski,C.,and F., Molinari(2020) Estimating the COVID-19 Infection Rate:Anatomy of an Inference Problem,Northwestern Univ. DP.\\

\nin McFadden, D. (1984): Econometric Analysis of Qualitative Response Models,in Handbook of Econometrics,Vol 2,1395-1457, Elsevier. \\

\nin McKendrick,A. (1926): Applications of Mathematics to Medical Problems, Proceedings of the Edinburgh Mathematical Society, 14, 9-130.\\

\nin Miller,J. and G. Judge (2015): "Information Recovery in a Dynamic Statistical Markov Model", Econometrics, Vol 3/2, 187-198. \\ 

\nin McRae, E. (1977): Estimation of Time Varying Markov Processes with Aggregate Data,Econometrica, 45,183-198. \\

\nin Nishiura,H. et al. (2020): Estimation of the Asymptomatic Ratio of Novel Coronavirus Infection(COVID-19),Int.J. Infect.Dis. \\

\nin Pitt, M., and N., Shephard (1999):Filtering via Simulation: Auxiliary Particle Filters, JASA,94,590-599.\\

\nin Sante Publique France (2020): Donnees Hospitalieres Relatives a l'Epidemie Covid-19.\\

\nin Schon, T., Gustafson,F and P.J., Nordlund (2005): Marginalized Particle Filters for Mixed Linear/Nonlinear State Space Models, IEEE Trans. on Signal Processing, 53,2279-2288.\\

\nin Song,Y and J, Grizzle (1995): The Extended Kalman Filter as a Local Asymptotic Observer,Estimation and Control,5,59-78 \\

\nin Verity, R. et al. (2019): Estimates of the Severity of Coronavirus Disease 2019;A Model Based Analysis,Lancet Infect.Dis.\\

\nin Vynnicky,.E., and White,R. (eds)(2010): An Introduction to Infectious Disease Modelling, Oxford Univ Press. \\

\nin Tang,B et al.(2020) : Estimation of the transmission Risk of the 2019-nCoV and its Implication for Public Health Interventions,Journal of Clinical Medicine,9;\\
 
 Wan,E and R,Van der Merwe (2000) : The Unscented Kalman Filter for Nonlinear Estimation,in Adaptive Systems for Signal Processing,Communication and Control Symposium,IEEE 2000,153-158 \\

\nin Wu,K, Darcet,D,Wang,Q and D,Sornette (2020): Generalized Logistic Growth Modelling of the Covid 19 Outbreak in 29 Provinces in China and the Rest of the World,DP Univ.Zurich.
 
\newpage

\bc Appendix 1 \\

{\bf Expression of the Autocovariance Operator}
\ec

Instead of characterizing the individual histories  by the qualitative sequences $Y_{it}$, a sequence of J-dimensional vectors $Z_{it}$ can be alternatively considered, where component $j$ is the 0-1 indicator of $Y_{it}=j$. Then we have:

$$ E( Z_t| Z_{t-1})= P(t-1) Z_{t-1},$$

\nin where $P(t-1)$ denotes the transition matrix from date t-1 to date t. By the iterated expectations theorem, we get:

$$ E(Z_t| Z_{t-h})= \Pi(t-1;h) Z_{t-h},$$

\nin where $\Pi(t-1;h)= P(t-1)...P(t-h)$.

\medskip

Let us now consider the covariance:

\begin{eqnarray*}
\Omega_{t,t-h} & = & Cov(Z_t,Z_{t-h}) = E(Z_t Z_{t-h}')- E(Z_t) E(Z_{t-h})'\\
& = & E(\Pi(t-1;h) Z_{t-h} Z_{t-h}') -p(t) p(t-h)'
\end{eqnarray*}

\nin (by the iterated expectation and using $E(Z_t)= p(t)$)

$$ = \Pi(t-1;h) E[ diag(Z_{t-h})] - p(t) p(t-h)'$$

\nin (by taking into account the 0-1 components of Z)

$$ = \Pi(t-1;h) diag[p(t-h)] - p(t) p(t-h)'.$$

\nin This is the expression of the autocovariance as a function of the $p(t)$'s and model parameters.
 
\newpage

\bc
Appendix 2 \\

{\bf Asymptotic Expansions}
\ec

They are easily derived, given that the optimization in Proposition 2 is deterministic. Therefore, estimators $\hat{p}(1), \hat{p}(2),...,\hat{p}(T), \hat{\theta}$ are deterministic functions of observations $\hat{A}_t = A f(t),t=1,..,T$. If the transition matrix is twice continuously differentiable with respect to $p(t-1)$ and $\theta$ in a neighbourhood of the true values, these deterministic functions are continuously differentiable. Then, by using the asymptotic normality of $f(t)$'s (Proposition 1), we can apply the delta method to deduce the $1/\sqrt{N}$ rate of convergence of the estimators and their asymptotic variance-covariance matrix from the one of the $f(t)$'s (see Appendix 1).

When the number of observation dates and of missing counts is too large, the use of the delta method can be numerically cumbersome. It can be replaced by a bootstrap method (for which the regularity conditions of validity are satisfied in our framework, or by the approximated accuracies provided by an EKF, or UKF algorithm.

\bc
Appendix 3 \\

{\bf Nongeneric Cases in Proposition 3}
\ec

This appendix derives the equations used in the proof of Proposition 3. It provides the closed form expressions of functions $a(P), b(P), c(P)$, and outlines conditions 1 to 4 for the validity of Proposition 3.

\medskip

i) Let us first solve the second equation of system (4.3). We get:

$$ (p_{23}- p_{13}) p_2(t-1) = p_3(t) + (p_{13}- p_{33}) p_3(t-1 ) - p_{13},$$

\nin  or,

$$ p_2(t-1)= [p_3(t) + (p_{13}-p_{33}) p_3(t-1) -p_{13}]/ (p_{23}-p_{13}),$$

\nin if the following condition is satisfied:

\medskip
\nin {\bf condition 1:} $p_{23}$ is different of $p_{13}$.

\medskip
ii) Next, let us consider the first equation of system (4.3):

$$ p_2(t)= p_{12} + (p_{22}-p_{12})p_2(t-1) + (p_{32}-p_{12}) p_3(t-1) $$

\nin and substitute into this equation the expression of $p_2(t)$ derived in part i). We get:

$p_3(t+1) + ( p_{13}-p_{33}) p_3(t) - p_{13}
= p_{12} (p_{23}-p_{13}) + (p_{22}-p_{12}) [ p_3(t) + (p_{13}-p_{33}) p_3(t-1) -p_{13}] +(p_{23}-p_{13}) (p_{32}-p_{12}) p_3(t-1).$

\medskip

\nin It follows that:

\vspace{-0.1in}
\begin{eqnarray*}
a(P)& =& p_{12}( p_{23}-p_{13}) + p_{13} (1-p_{22}+p_{12}), \\
b(P)&= & p_{22}-p_{12} +p_{33}-p_{13}, \\
c(P) &=& (p_{22} -p_{12})(p_{13}-p_{33}) + (p_{23}-p_{13}) (p_{32}-p_{12}).
\end{eqnarray*}

\nin To get a recursive equation of order 2, we need the second condition:

\medskip
\nin {\bf condition \ 2:} $c(P) \neq 0$

\medskip

\nin To identify functions $a,b,c$ from the observed $p_3(t)$, we need:

\medskip

\nin {\bf condition 3:} The matrix $3 \times (T-2)$ with columns $(1,...,1)', (p_3(T-1),p_3(T-2),..,p_3 (2))'$ and $(p_3(T-2), p_3(T-3),...,p_3(1))'$ is of full column rank.
\medskip

\nin This implies, in particular, the order condition: $T\geq 5 $  in Proposition 3.

\medskip
\nin The following condition 4 is needed for computing the exact under-identification order of $P$ from functions $a,b,c$.
\medskip

\nin {\bf condition 4:} By taking into account the unit mass restrictions on the rows of $P$,the Jacobian of $(a,b,c)$ has rank 3.

\medskip
\nin Note that condition 4 implies condition 2.

\begin{figure*}[t!] 
\centering 
  \begin{subfigure}[t]{0.4\textwidth}
  \centering
  {\captionsetup{position=top}
            \phantomsubcaption{$p_2(t)$}
    \includegraphics[width=\linewidth]{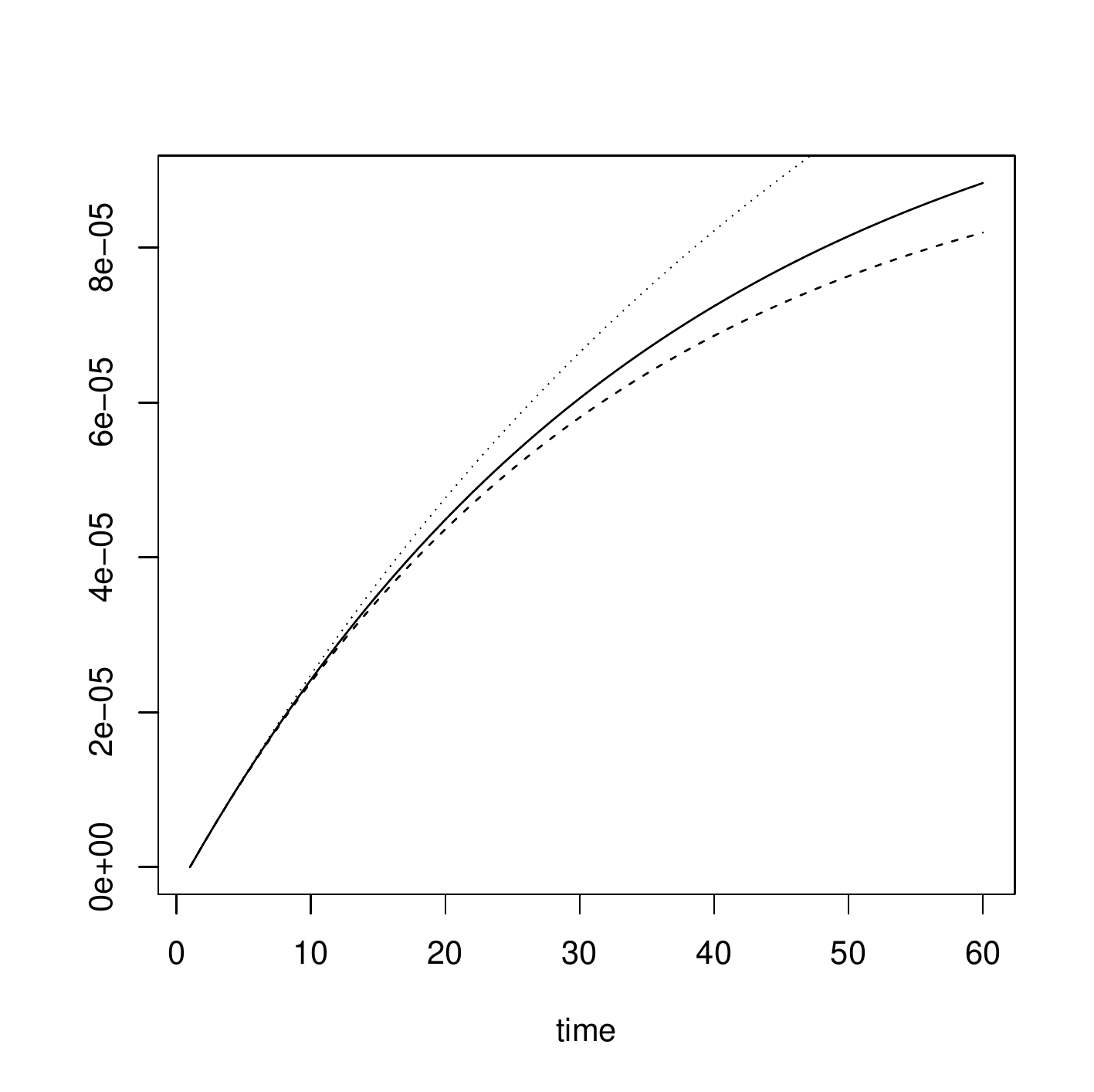}}
  \end{subfigure}
   \begin{subfigure}[t]{0.4\textwidth}
  \centering
  {\captionsetup{position=top}
            \phantomsubcaption{$p_3(t)$}
    \includegraphics[width=\linewidth]{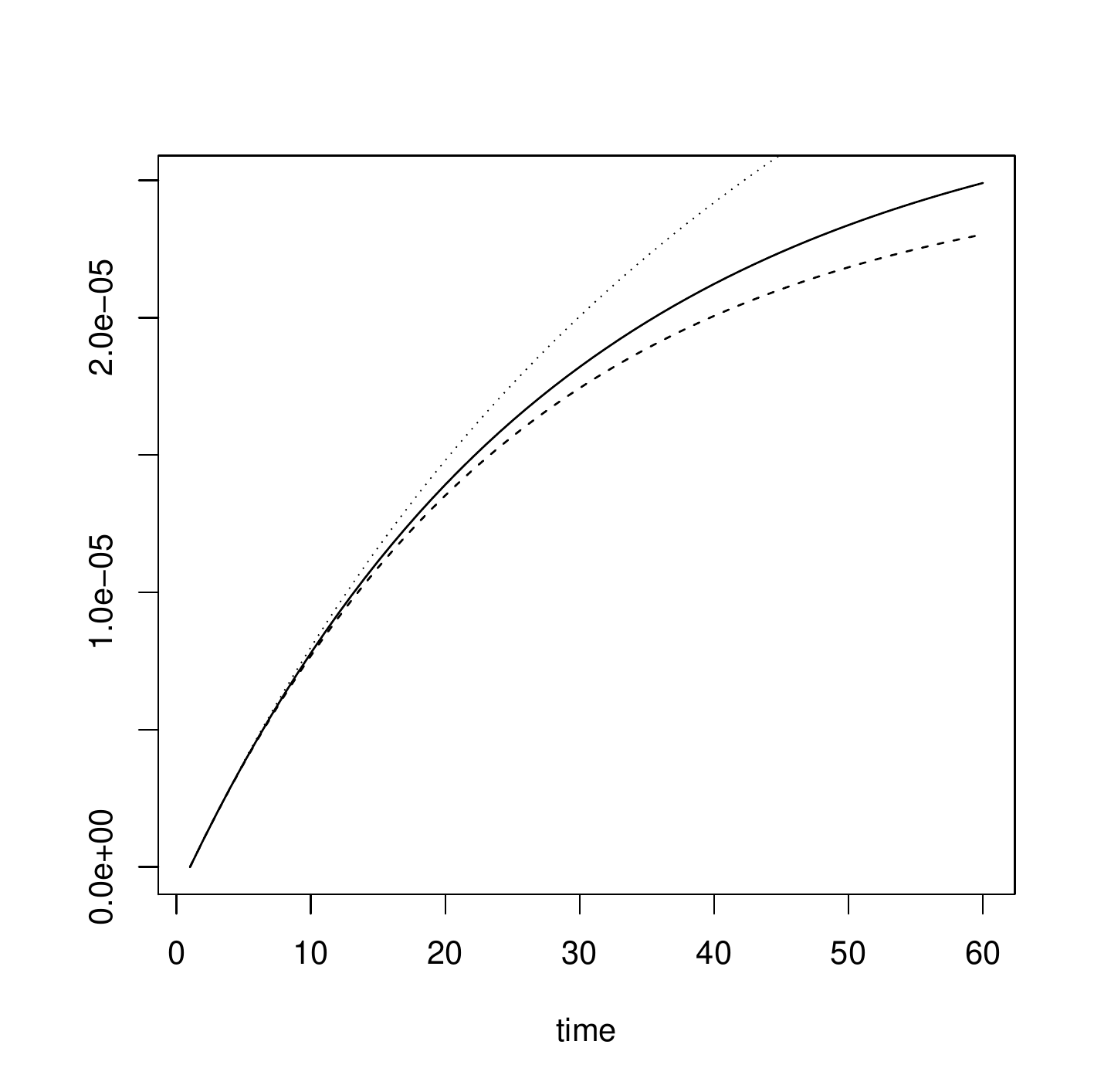}}
  \end{subfigure}
    \begin{subfigure}[t]{0.4\textwidth}
  \centering
  {\captionsetup{position=top}
            \phantomsubcaption{$p_4(t)$}
    \includegraphics[width=\linewidth]{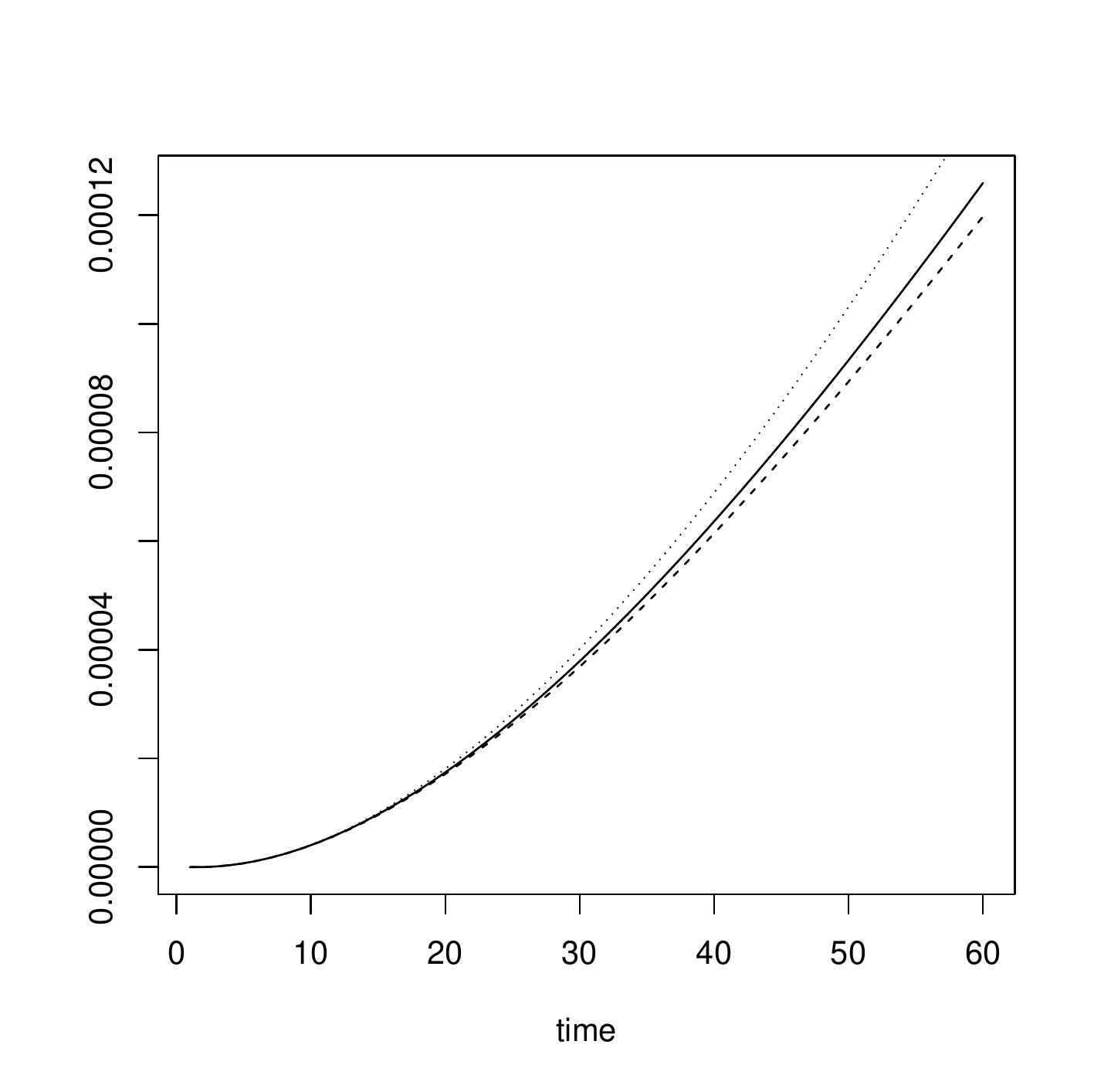}}
  \end{subfigure}
  \begin{subfigure}[t]{0.4\textwidth}
  \centering
  {\captionsetup{position=top}
            \phantomsubcaption{$p_5(t)$}
    \includegraphics[width=\linewidth]{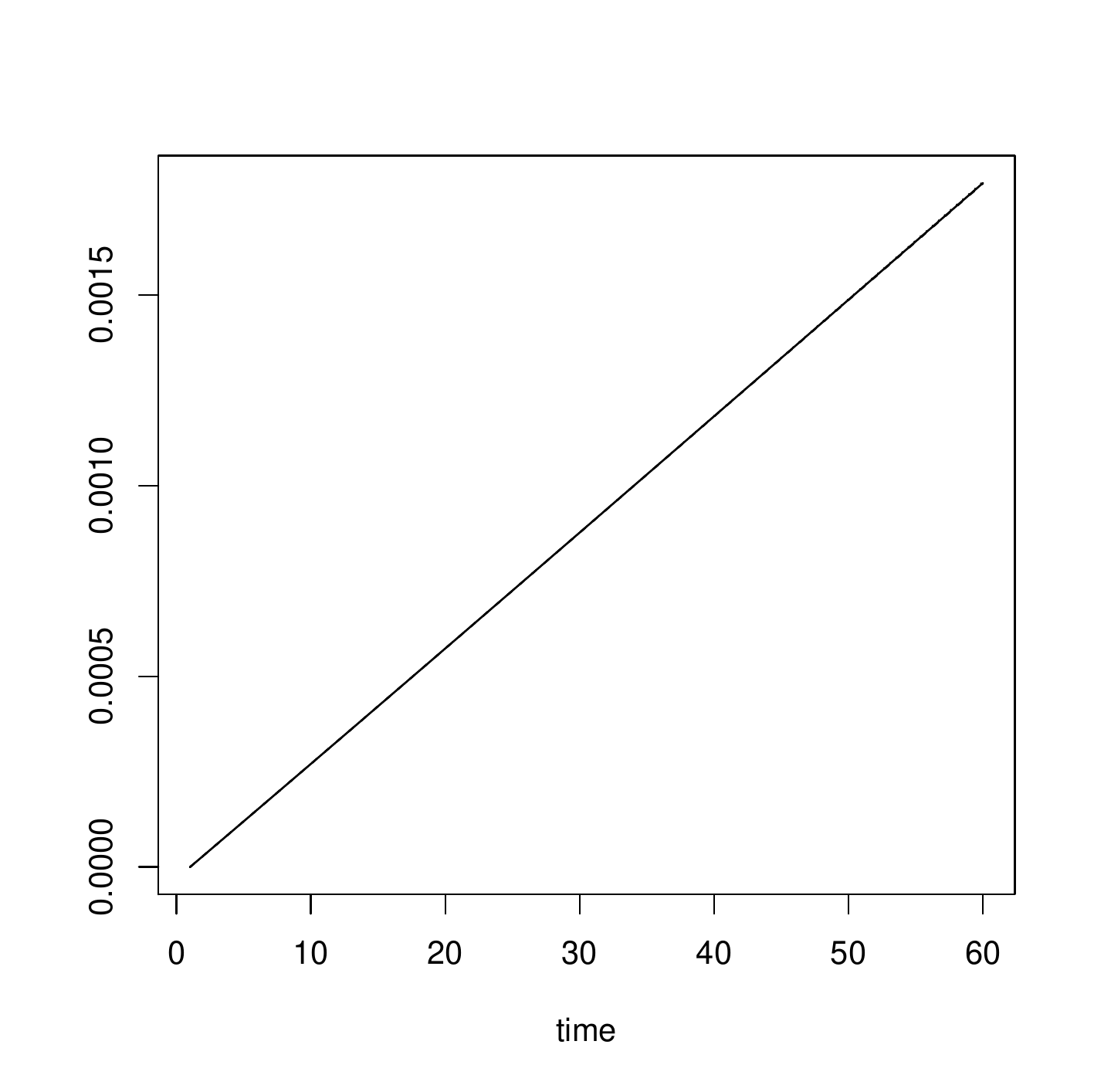}}
  \end{subfigure}
 \medskip

  Figure 1: Evolution of Marginal Probabilities \\
  Solid line-baseline, dotted line - doubled propagation parameters, dashed line - halved propagation parameters
\end{figure*}

\begin{figure*}[t!] 
\centering 
  \begin{subfigure}[t]{0.4\textwidth}
  \centering
  {\captionsetup{position=top}
            \phantomsubcaption{$\Delta p_3(t)*pop$}
    \includegraphics[width=\linewidth]{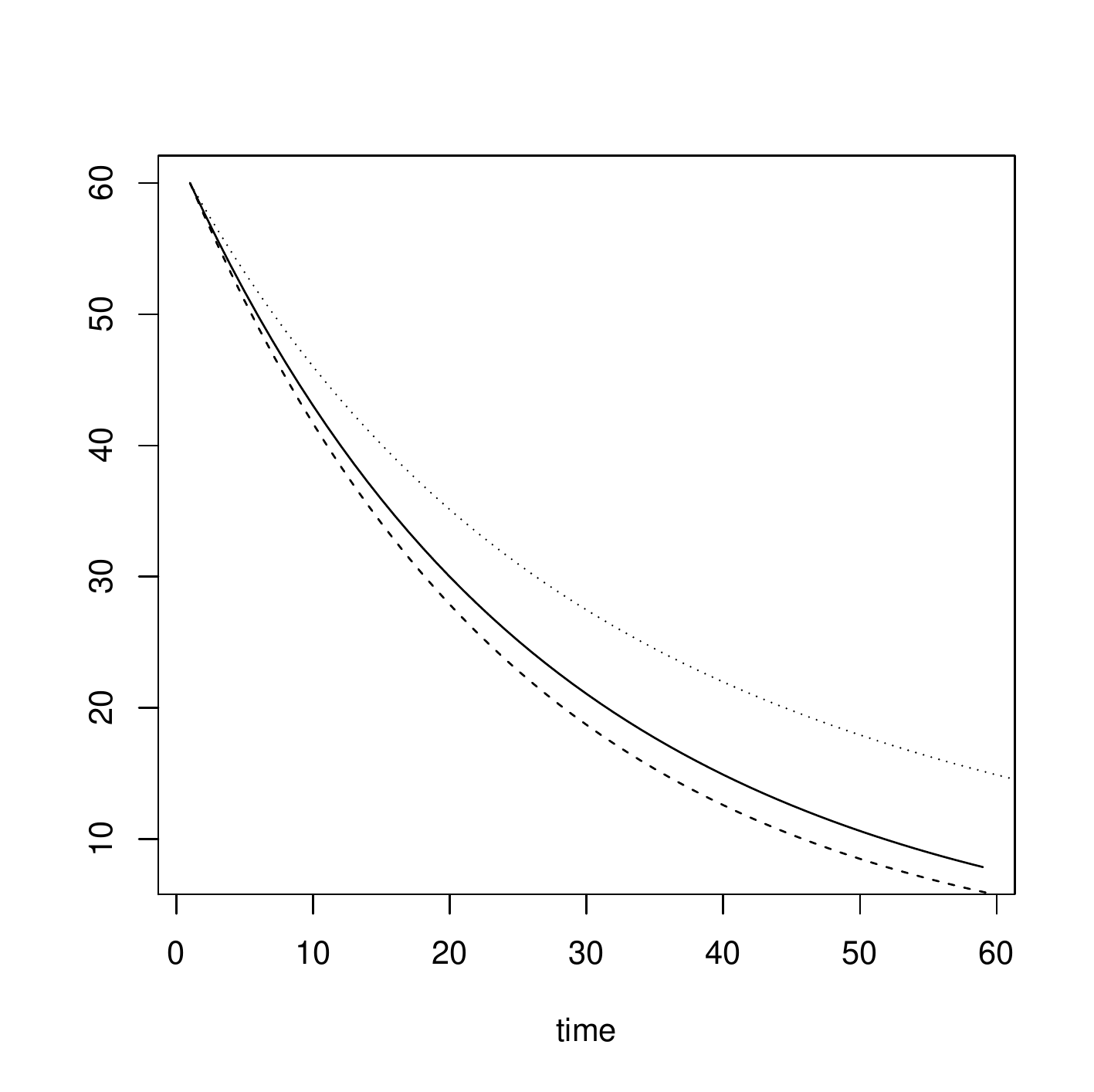}}
    \end{subfigure}
   \begin{subfigure}[t]{0.4\textwidth}
  \centering
  {\captionsetup{position=top}
            \phantomsubcaption{$\Delta p_5(t)*pop$}
    \includegraphics[width=\linewidth]{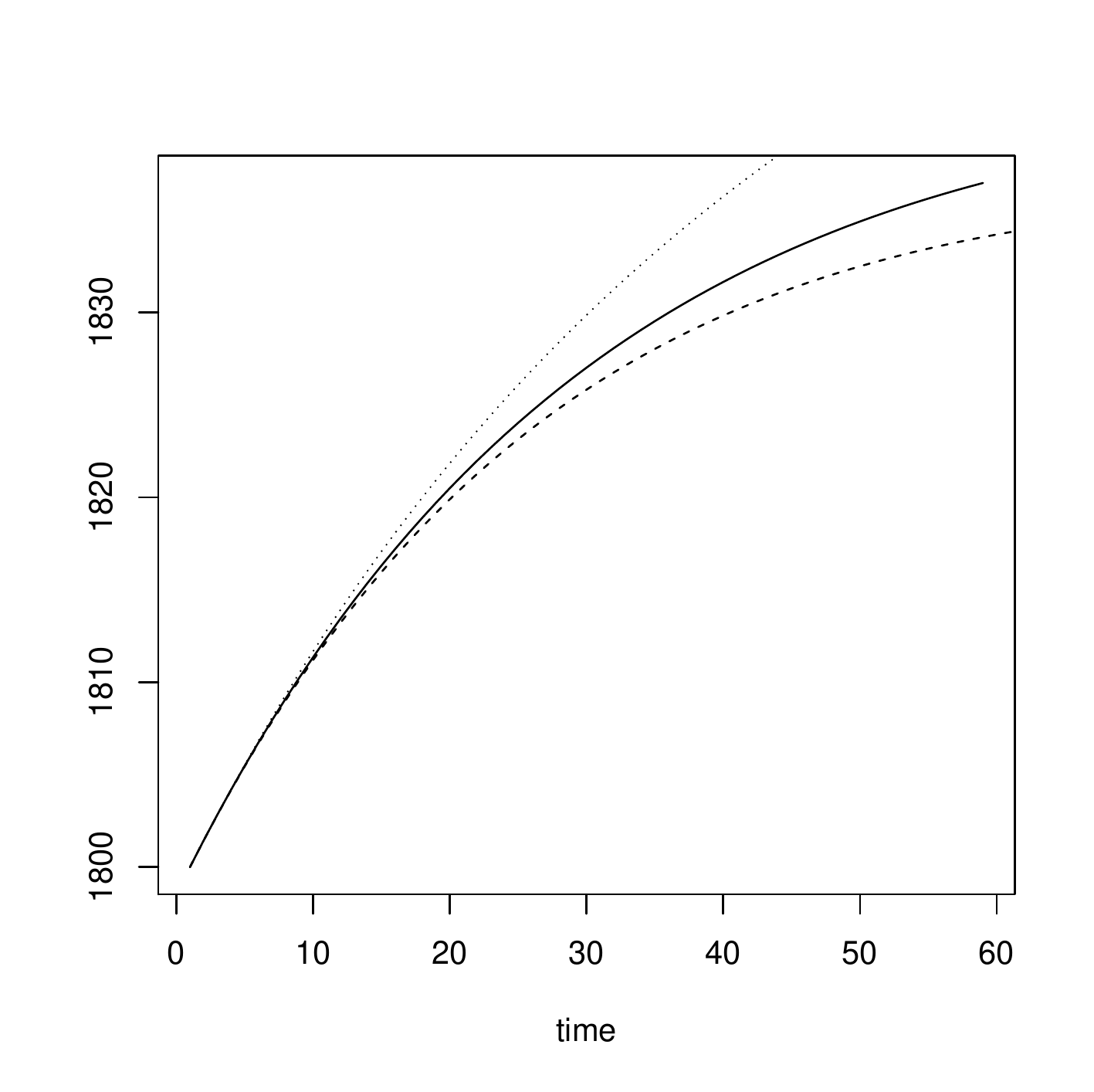}}
  \end{subfigure} 
   \medskip

  Figure 2: Evolution of New Counts \\
  Solid line-baseline, dotted line - doubled propagation parameters, dashed line - halved propagation parameters
\end{figure*}

\begin{figure*}[t!] 
\centering 
  \begin{subfigure}[t]{0.4\textwidth}
  \centering
  {\captionsetup{position=top}
            \phantomsubcaption{hospitalized}
    \includegraphics[width=\linewidth]{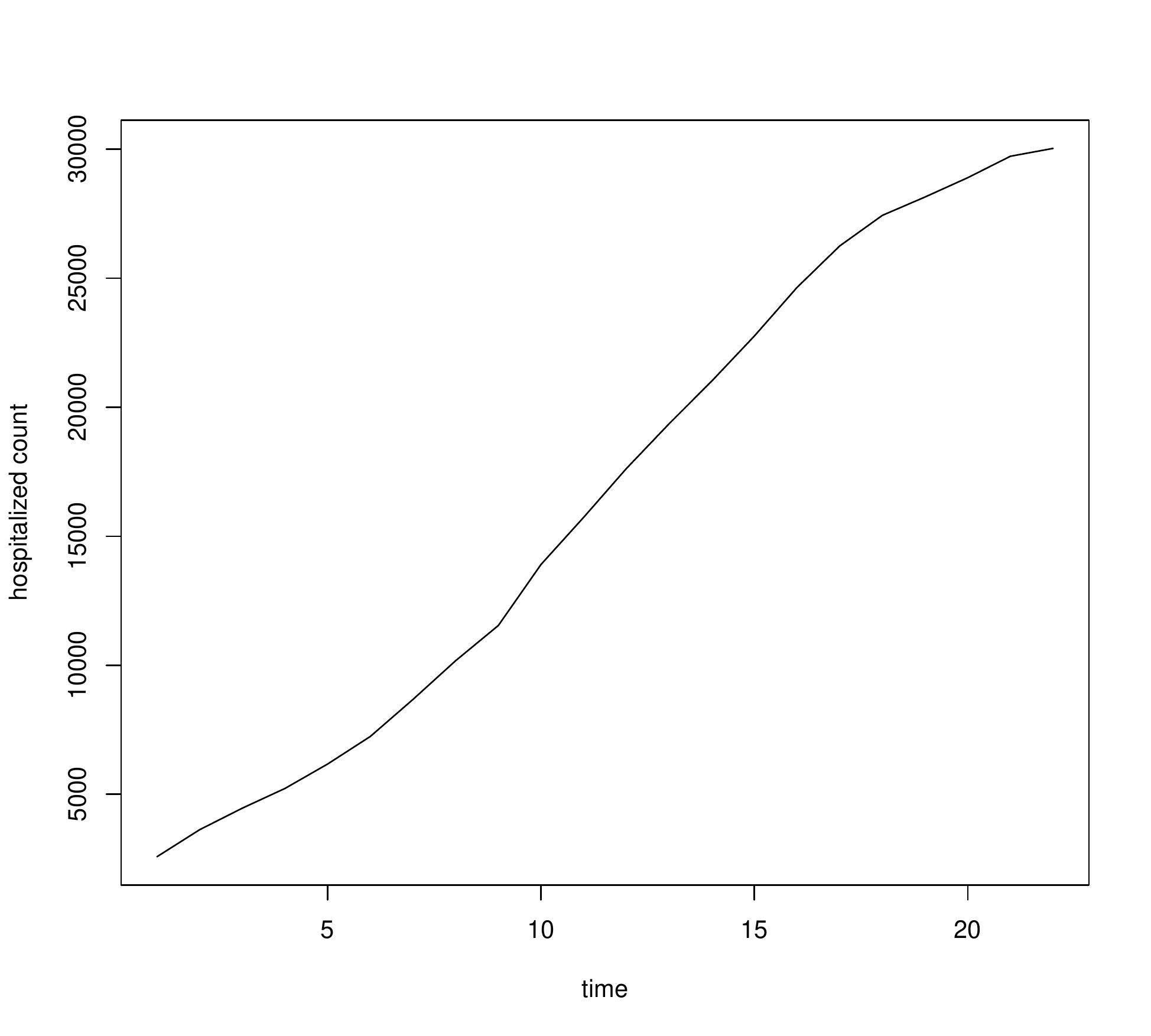}}
  \end{subfigure}
   \begin{subfigure}[t]{0.4\textwidth}
  \centering
  {\captionsetup{position=top}
            \phantomsubcaption{detected}
    \includegraphics[width=\linewidth]{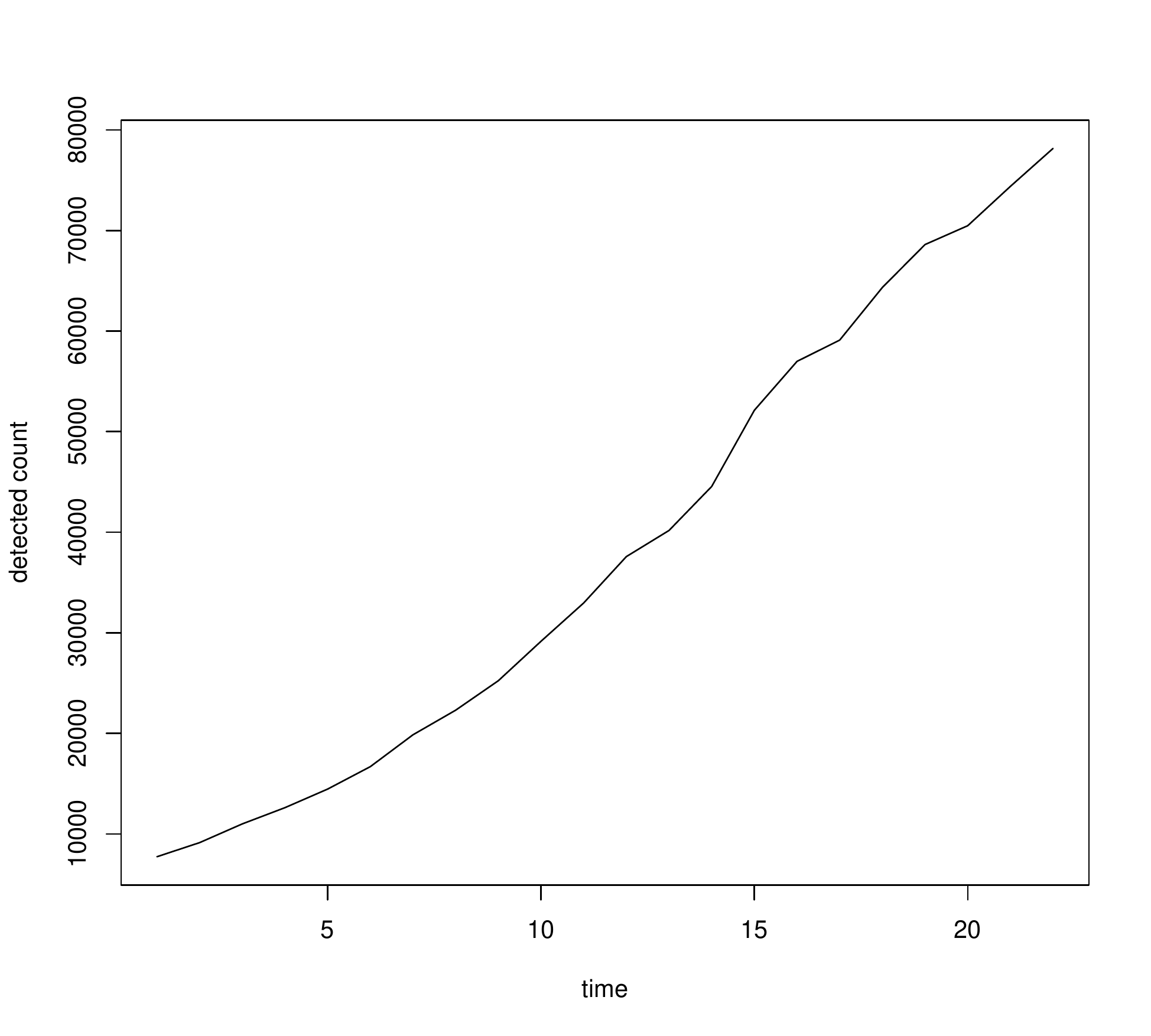}}
  \end{subfigure}
    \begin{subfigure}[t]{0.39\textwidth}
  \centering
  {\captionsetup{position=top}
            \phantomsubcaption{recovered}
    \includegraphics[width=\linewidth]{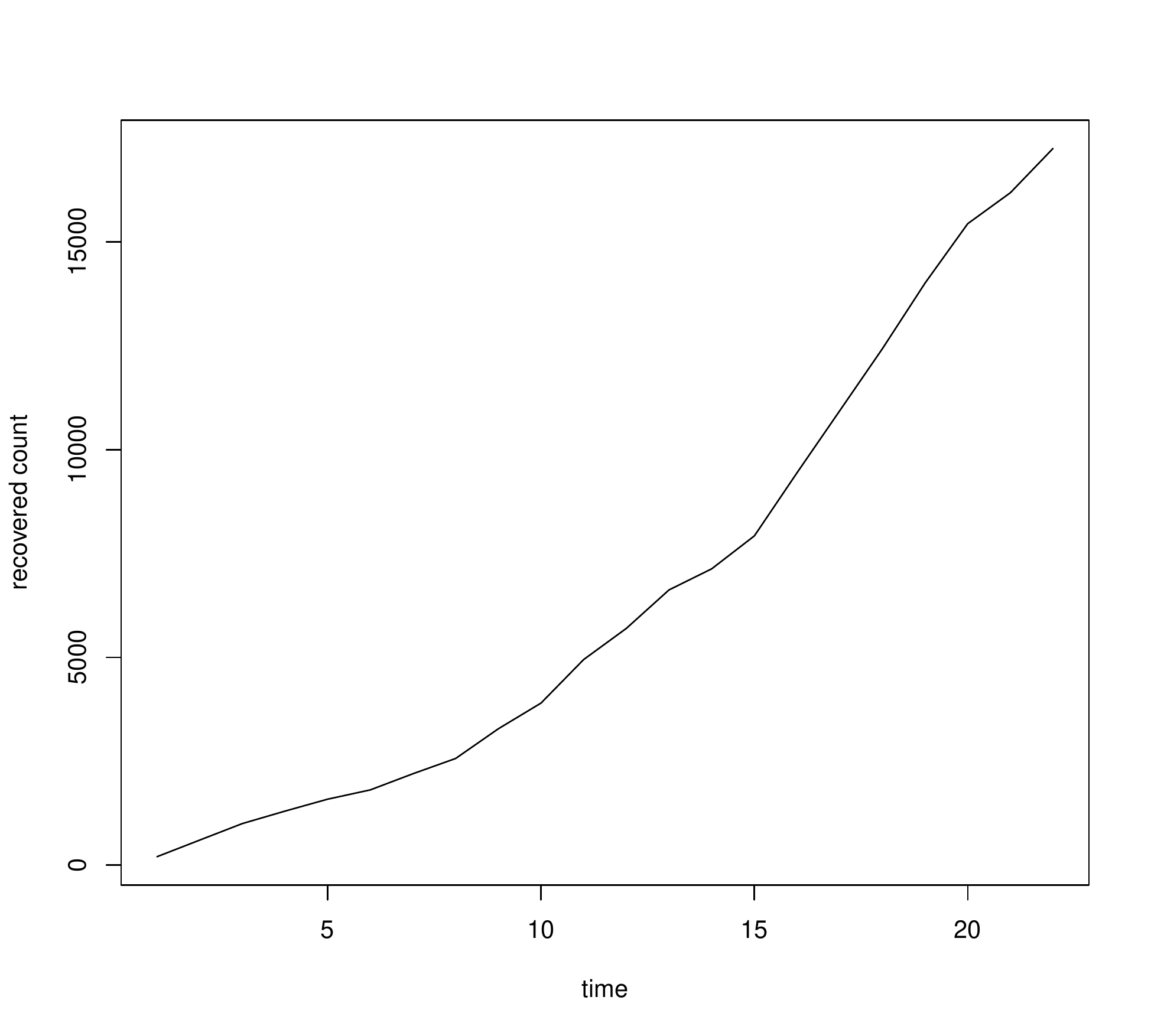}}
  \end{subfigure}
  \begin{subfigure}[t]{0.39\textwidth}
  \centering
  {\captionsetup{position=top}
            \phantomsubcaption{deceased}
    \includegraphics[width=\linewidth]{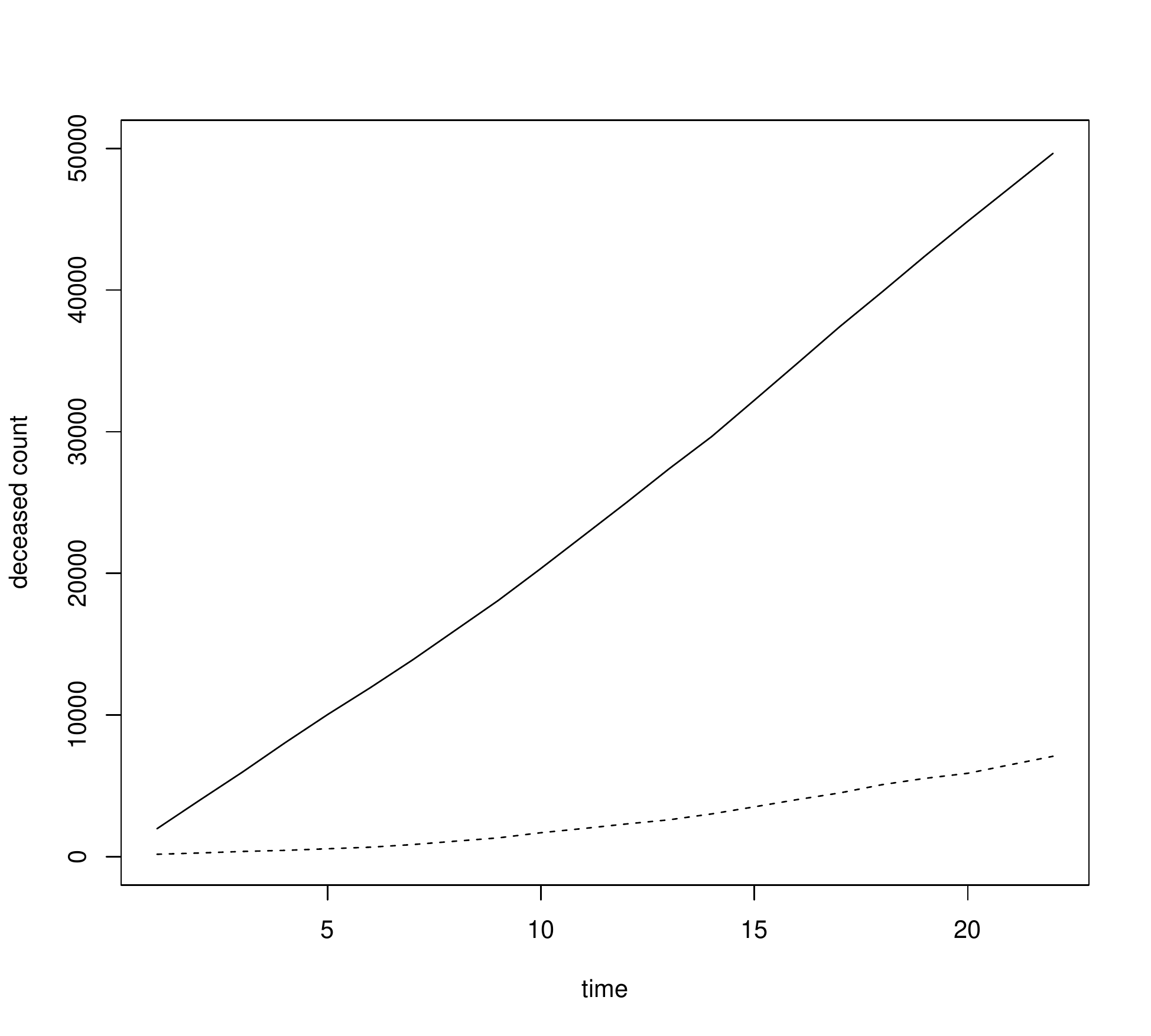}}
  \end{subfigure}
 \medskip

  Figure 3: Evolution of Observed Counts, 03/16 to 04/06, France \\
  The figure shows the evolution of observed daily counts. In the panel of deceased (bottom, right), the solid line shows the total deceased in France and the dashed line the (reported) deceased due to Covid-19
\end{figure*}

\begin{figure*}[t!] 
\centering 
  \begin{subfigure}[t]{0.65\textwidth}
  \centering
  {\captionsetup{position=top}
            \phantomsubcaption{estimated IU and observed ID}
    \includegraphics[width=\linewidth]{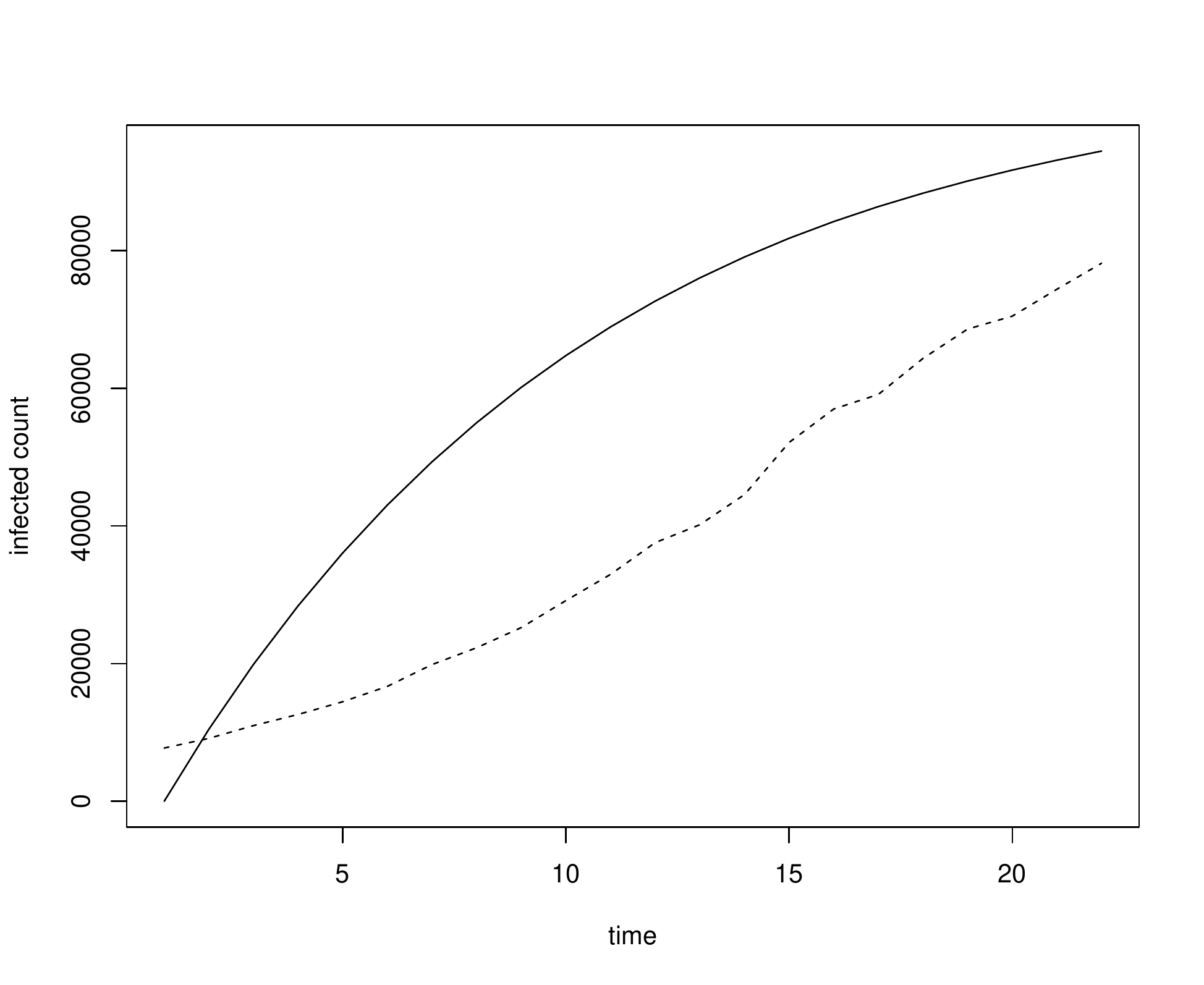}}
    \end{subfigure}
   \begin{subfigure}[t]{0.6\textwidth}
  \centering
  {\captionsetup{position=top}
            \phantomsubcaption{estimated and observed Recovered}
    \includegraphics[width=\linewidth]{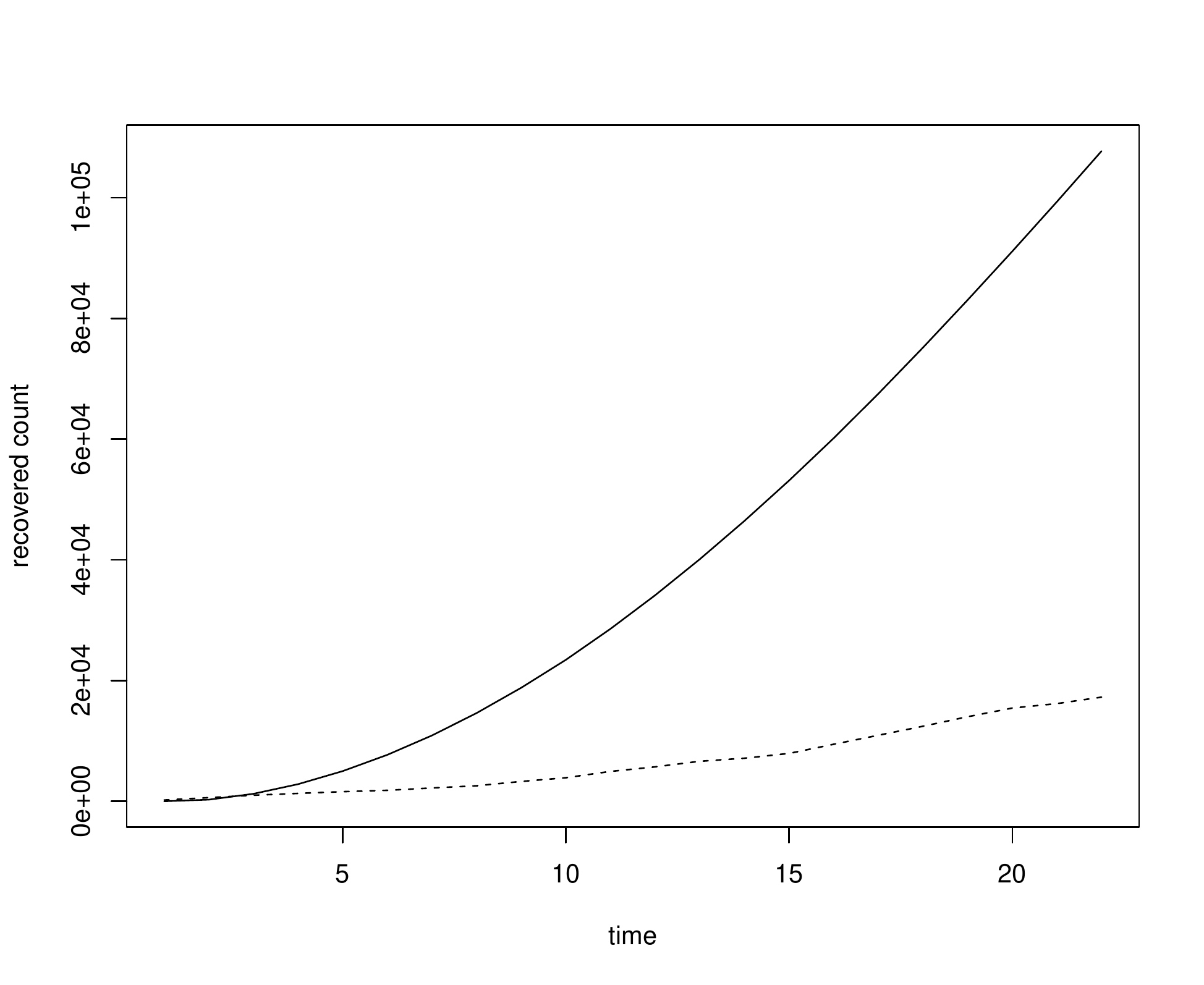}}
  \end{subfigure} 
   \medskip

  Figure 4: Estimated and Observed Counts \\
  The estimated counts - solid line, observed counts - dashed line. The figure compares the estimated counts of Infected and Undetected with the observed Infected and Detected (top panel), and Recovered estimated and reported R|H counts (bottom panel). 
\end{figure*}

\begin{figure*}[t!] 
\centering 
  \begin{subfigure}[t]{0.4\textwidth}
  \centering
  {\captionsetup{position=top}
            \phantomsubcaption{projected $p_2(t)$}
    \includegraphics[width=\linewidth]{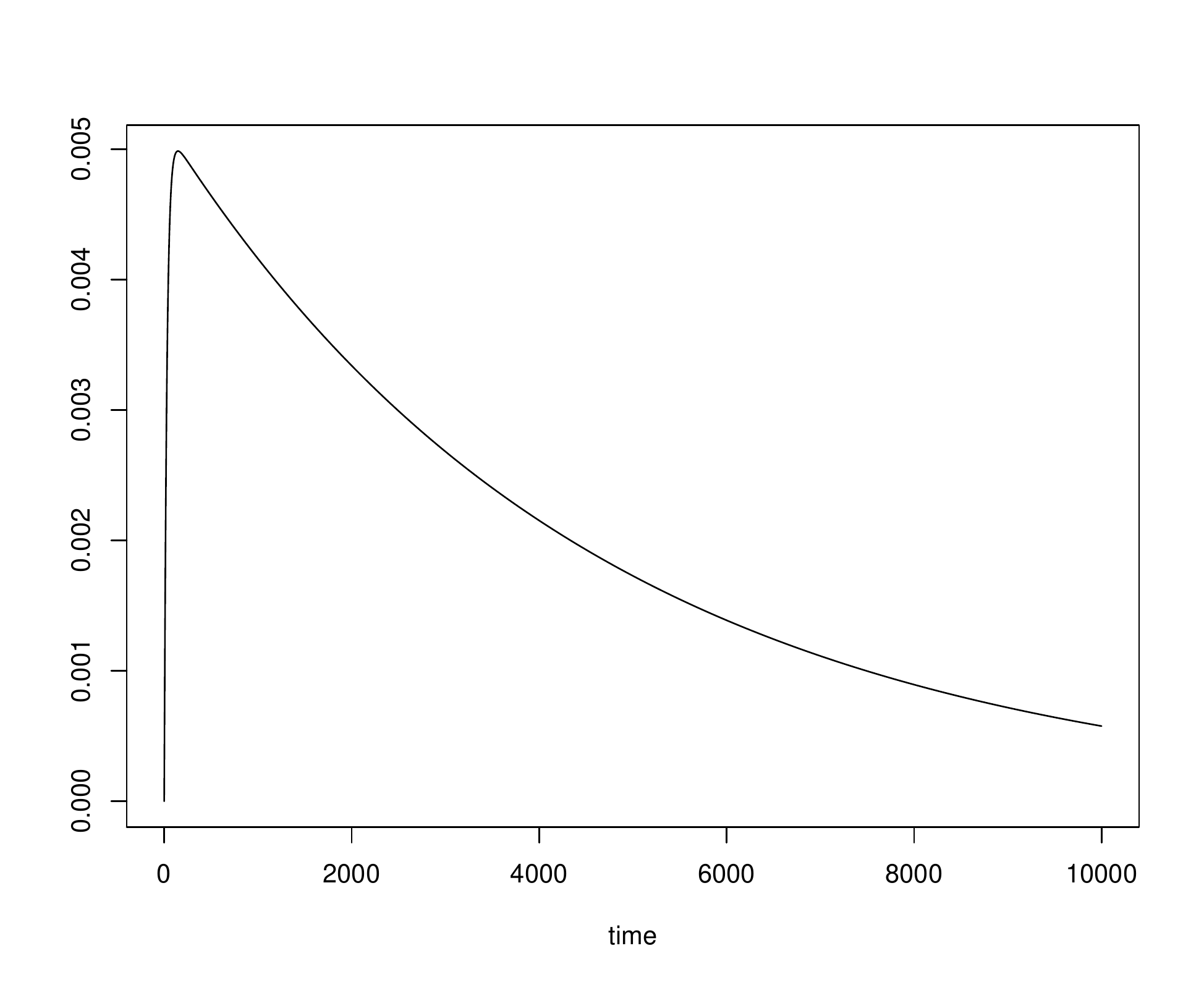}}
  \end{subfigure}
   \begin{subfigure}[t]{0.4\textwidth}
  \centering
  {\captionsetup{position=top}
            \phantomsubcaption{projected $p_3(t)$}
    \includegraphics[width=\linewidth]{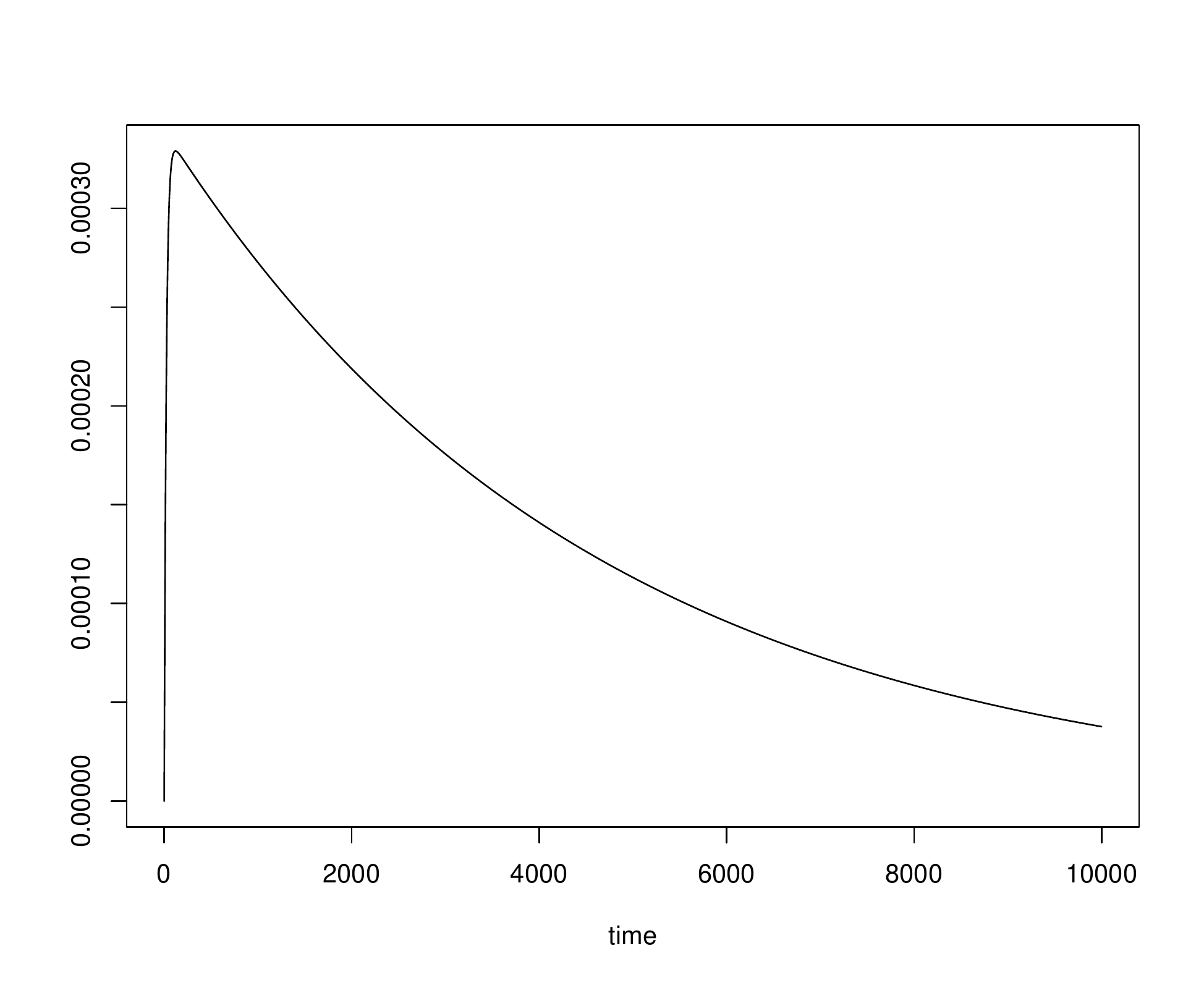}}
  \end{subfigure}
    \begin{subfigure}[t]{0.41\textwidth}
  \centering
  {\captionsetup{position=top}
            \phantomsubcaption{projected $p_4(t)$}
    \includegraphics[width=\linewidth]{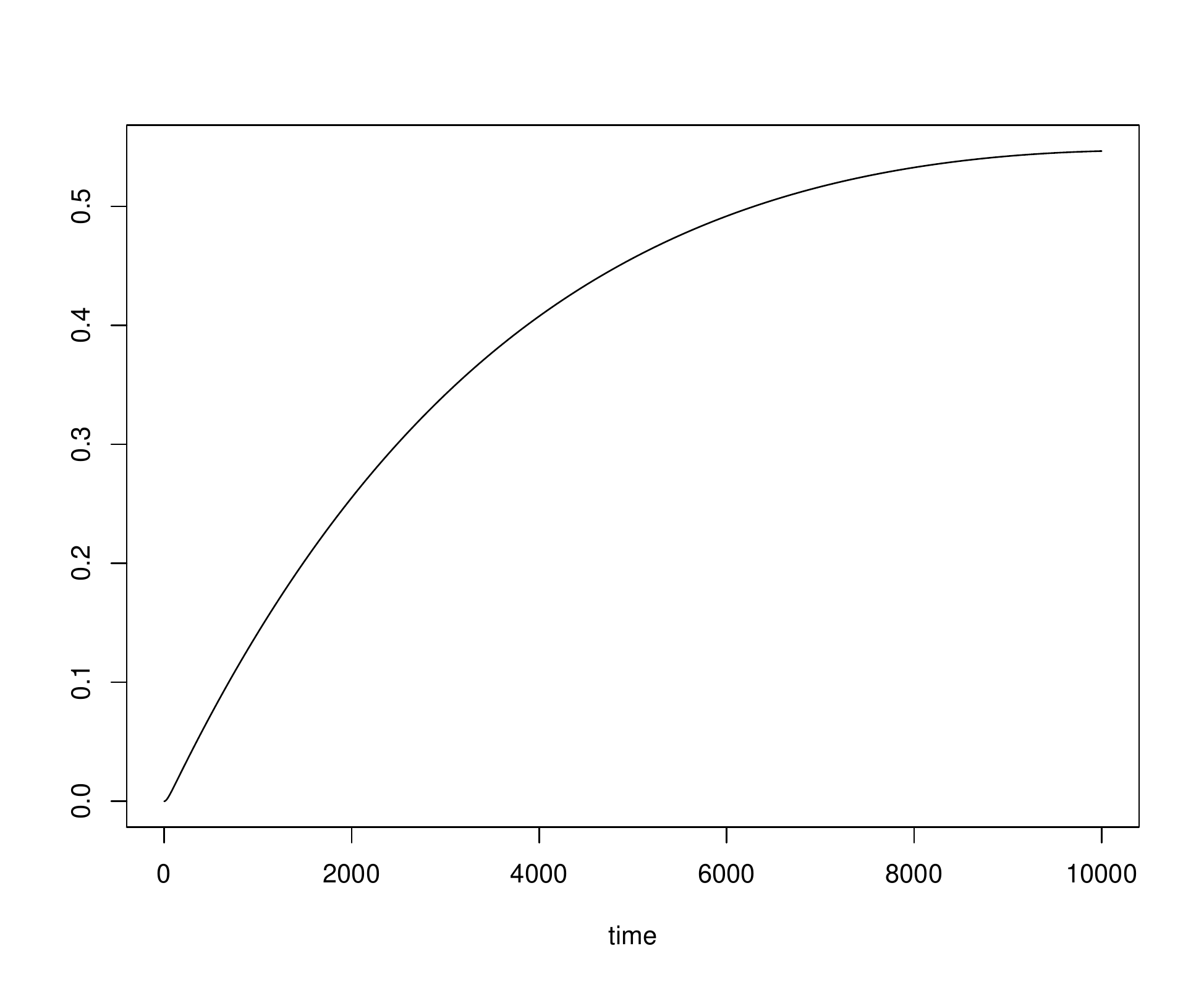}}
  \end{subfigure}
  \begin{subfigure}[t]{0.39\textwidth}
  \centering
  {\captionsetup{position=top}
            \phantomsubcaption{projected $p_5(t)$}
    \includegraphics[width=\linewidth]{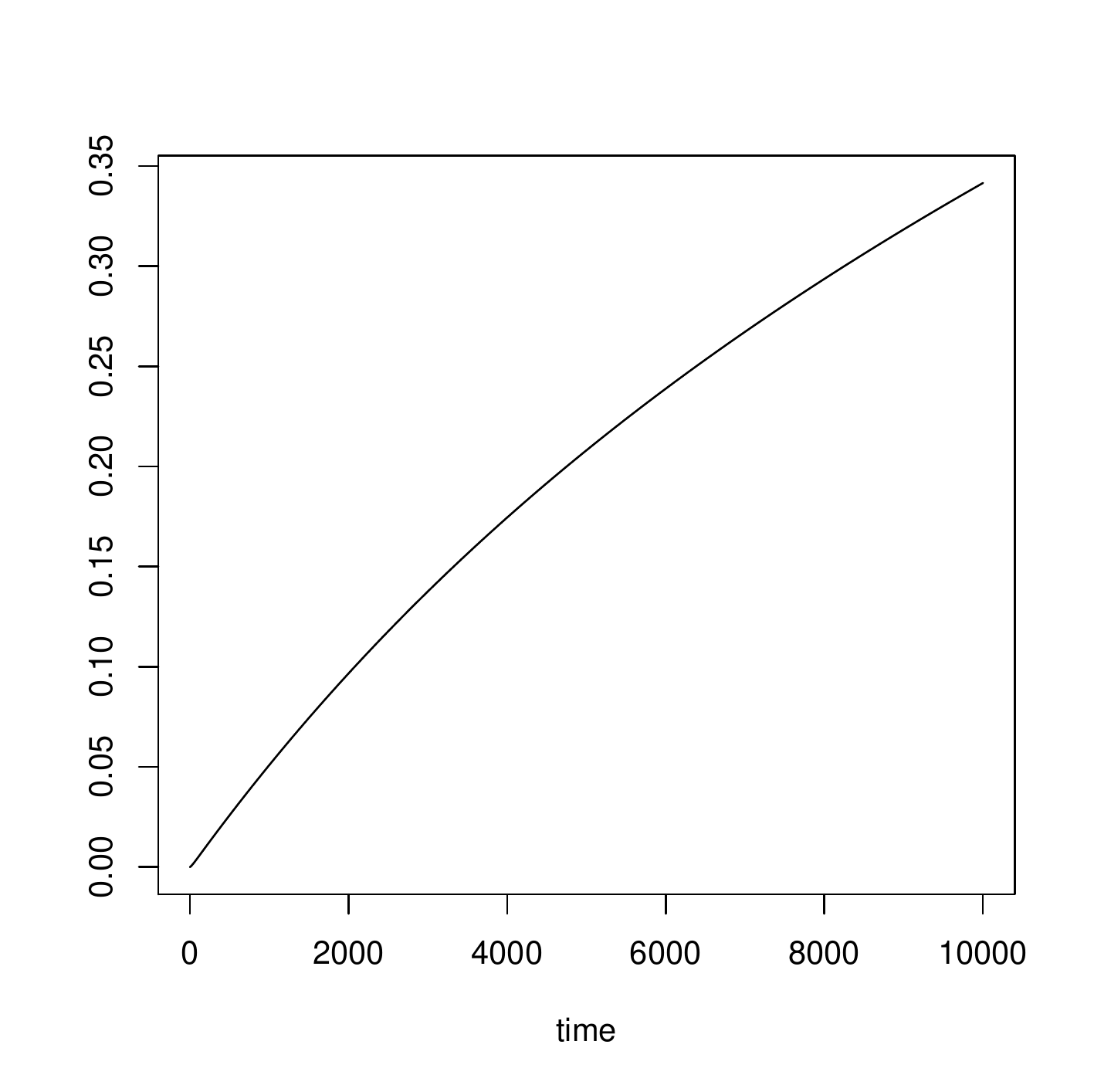}}
  \end{subfigure}
 \medskip

  Figure 5: Projected Evolution of Marginal Probabilities. \\
  
  The figure displays projected daily marginal probabilities of all states over 25 years.
\end{figure*}

\begin{figure*}[t!] 
\centering 
  \begin{subfigure}[t]{0.6\textwidth}
  \centering
  {\captionsetup{position=top}
            \phantomsubcaption{projected $\Delta p_3(t)*pop$}
    \includegraphics[width=\linewidth]{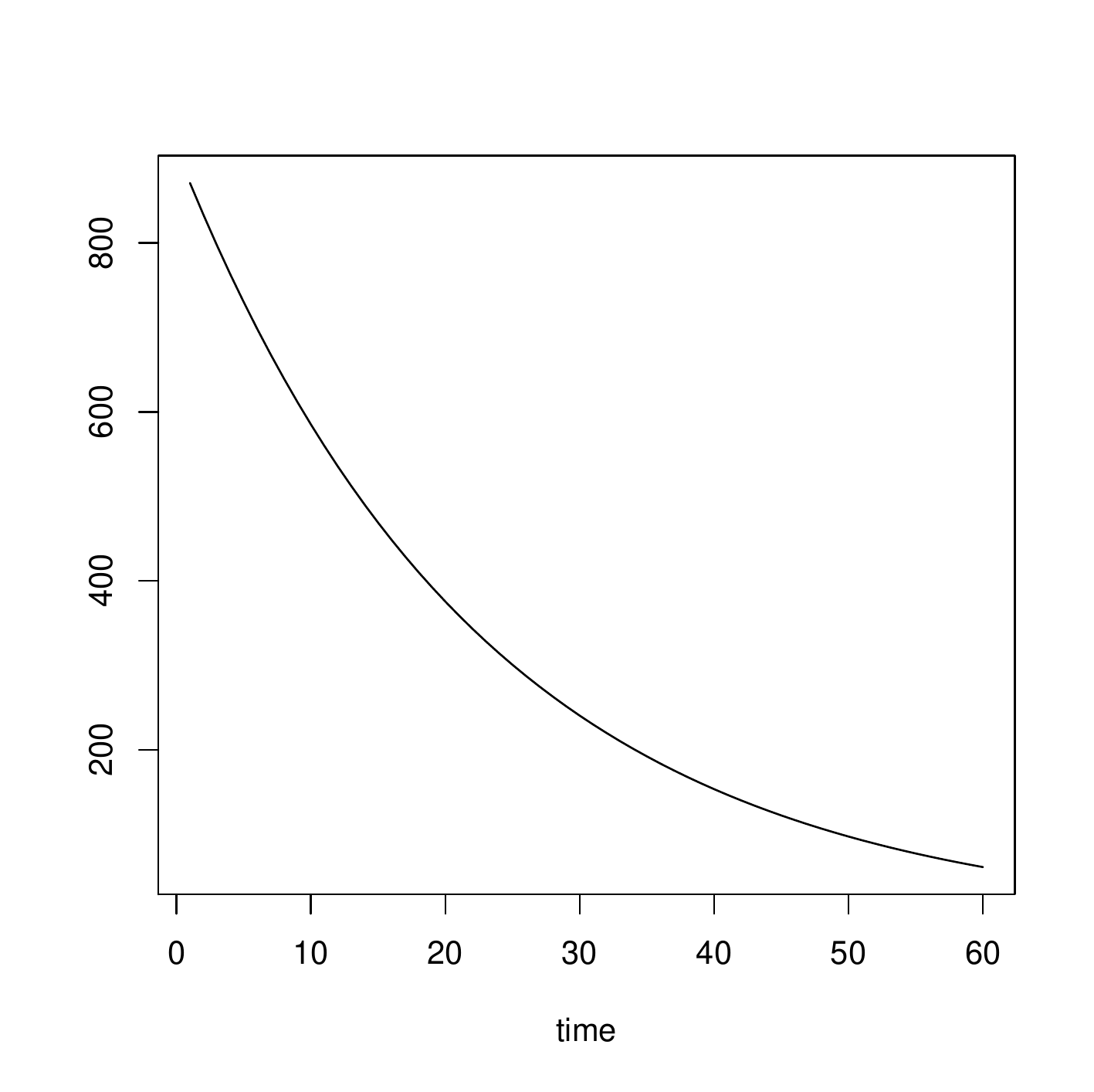}}
    \end{subfigure}
   \begin{subfigure}[t]{0.6\textwidth}
  \centering
  {\captionsetup{position=top}
            \phantomsubcaption{projected $\Delta p_5(t)*pop$}
    \includegraphics[width=\linewidth]{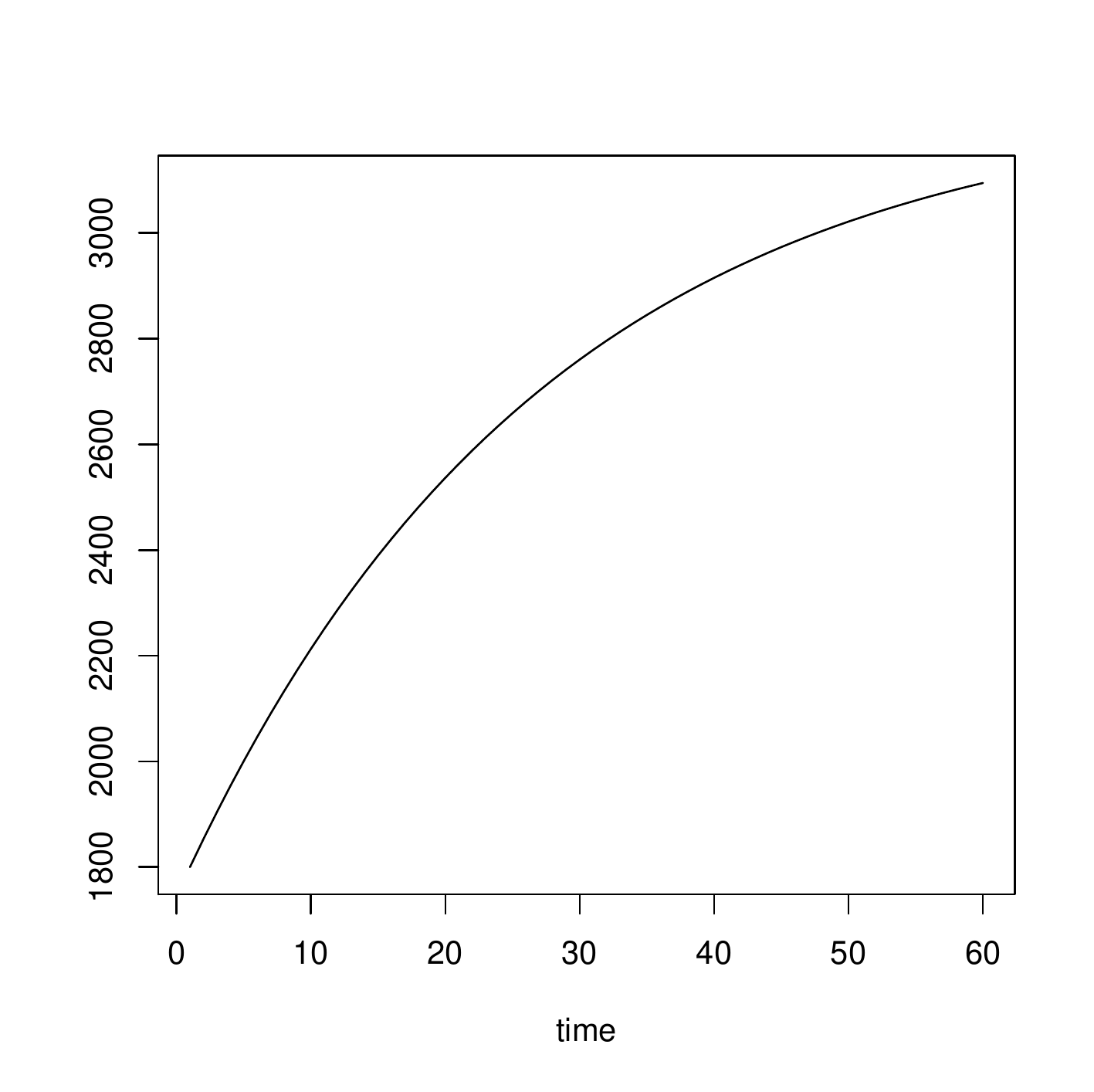}}
  \end{subfigure} 
   \medskip

  Figure 6: Projected Evolution of New Counts \\
  Figure 6 displays projected daily new counts in states 3 and 5 over 60 days
\end{figure*}

\end{document}